\documentclass[journal,comsoc]{IEEEtran}
\usepackage{graphicx}
\usepackage{subfigure}
\usepackage{float}
\usepackage{mathtools}
\usepackage{algorithm}
\usepackage{algorithmic}
\usepackage[normalem]{ulem}
\usepackage[tableposition=top]{caption}
\usepackage{url}
\usepackage{booktabs}

\usepackage{array}
\newcolumntype{?}[1]{!{\vrule width #1}}
\usepackage{color}
\usepackage{caption}
\usepackage{multirow}
\usepackage{url}
\usepackage{makecell}
\usepackage{dblfloatfix}

\usepackage[hidelinks]{hyperref}
\usepackage{amsthm}

\newtheoremstyle{mystyle}%                % Name
  {}%                                     % Space above
  {}%                                     % Space below
  {\itshape}%                                     % Body font
  {5pt}%                                     % Indent amount
  {\bfseries}%                            % Theorem head font
  {:}%                                    % Punctuation after theorem head
  { }%                                    % Space after theorem head, ' ', or \newline
  {}%                                     % Theorem head spec (can be left empty, meaning `normal')

\theoremstyle{mystyle}
\newtheorem{challenge}{Challenge}

\begin{document}

\title{Ubiquitous Acoustic Sensing on Commodity IoT Devices: A Survey}

\author{{~Chao~Cai,~\IEEEmembership{~Member,~IEEE,}
        ~Rong~Zheng,~\IEEEmembership{~Senior~Member,~IEEE,}
        ~Jun~Luo,~\IEEEmembership{~Senior~Member,~IEEE}
        }
\IEEEcompsocitemizethanks{ 
\IEEEcompsocthanksitem C. Cai is with the College of Life Science and Technology,
Huazhong University of Science and Technology. (E-mail: chriscai@hust.edu.cn)
\IEEEcompsocthanksitem R. Zheng is with the Department of Computing and Software, McMaster University, Canada (E-mail:rzheng@mcmaster.ca).
\IEEEcompsocthanksitem J. Luo is with the School of Computer Science and Engineering,
Nanyang Technological University, Singapore. (E-mail: junluo@ntu.edu.sg)x
}}

\markboth{IEEE Communication Surveys \& Tutorials,~Vol.~xx, No.~x, XX~2021}%
{Cai \MakeLowercase{\textit et al.}: Ubiquitous Acoustic Sensing on Commodity IoT Devices: A Survey}

\maketitle

\begin{abstract}
With the proliferation of Internet-of-Things devices, acoustic sensing attracts much attention in recent years. It  exploits acoustic transceivers such as microphones and speakers beyond their primary functions, namely recording and playing, to enable novel applications and new user experiences. 
In this paper, we present the first systematic survey of recent advances in active acoustic sensing using commodity hardware with a frequency range below 24~\!kHz. We propose a general framework that categorizes main building blocks of acoustic sensing systems. This framework encompasses three layers, i.e., physical layer, core technique layer, and application layer. The physical layer includes basic hardware components, acoustic platforms as well as the air-borne and structure-borne channel characteristics.
The core technique layer encompasses key mechanisms to generate acoustic signals (waveforms) and to extract useful temporal, spatial and spectral information from received signals. 
The application layer builds upon the functions offered by the core techniques to realize different acoustic sensing applications. 
We highlight unique challenges due to the limitations of physical devices and acoustic channels and how they are mitigated or overcame by core processing techniques and application-specific solutions. Finally, research opportunities and future directions are discussed to spawn further in-depth investigation on acoustic sensing.  
%Each successful application strives to handle lower-layer challenges specific to it, yet future research issues remain as existing challenges have not been fully addressed and new challenges are created by emerging application requirements. 
% After an in-depth analysis on the three-layer framework, we discuss potential challenges and future research trends in the end.
%
% We highlight different sensing approaches in the processing layer and fundamental design considerations in the physical layer. Many existing and potential applications including context-aware applications, human-computer interface, and aerial acoustic communications are presented in depth. Challenges and future research trends are also discussed.
\end{abstract}


\section{Introduction}
\label{sec:introduction}

%Recent years have witnessed a surge of Internet-of-Things (IoT)~\cite{IoTConcept} devices. The mobile devices, including smartphone, wearable devices, etc., contributes to the majority

\begin{figure}[!t]
  \centering
  \includegraphics[width=.89\columnwidth]{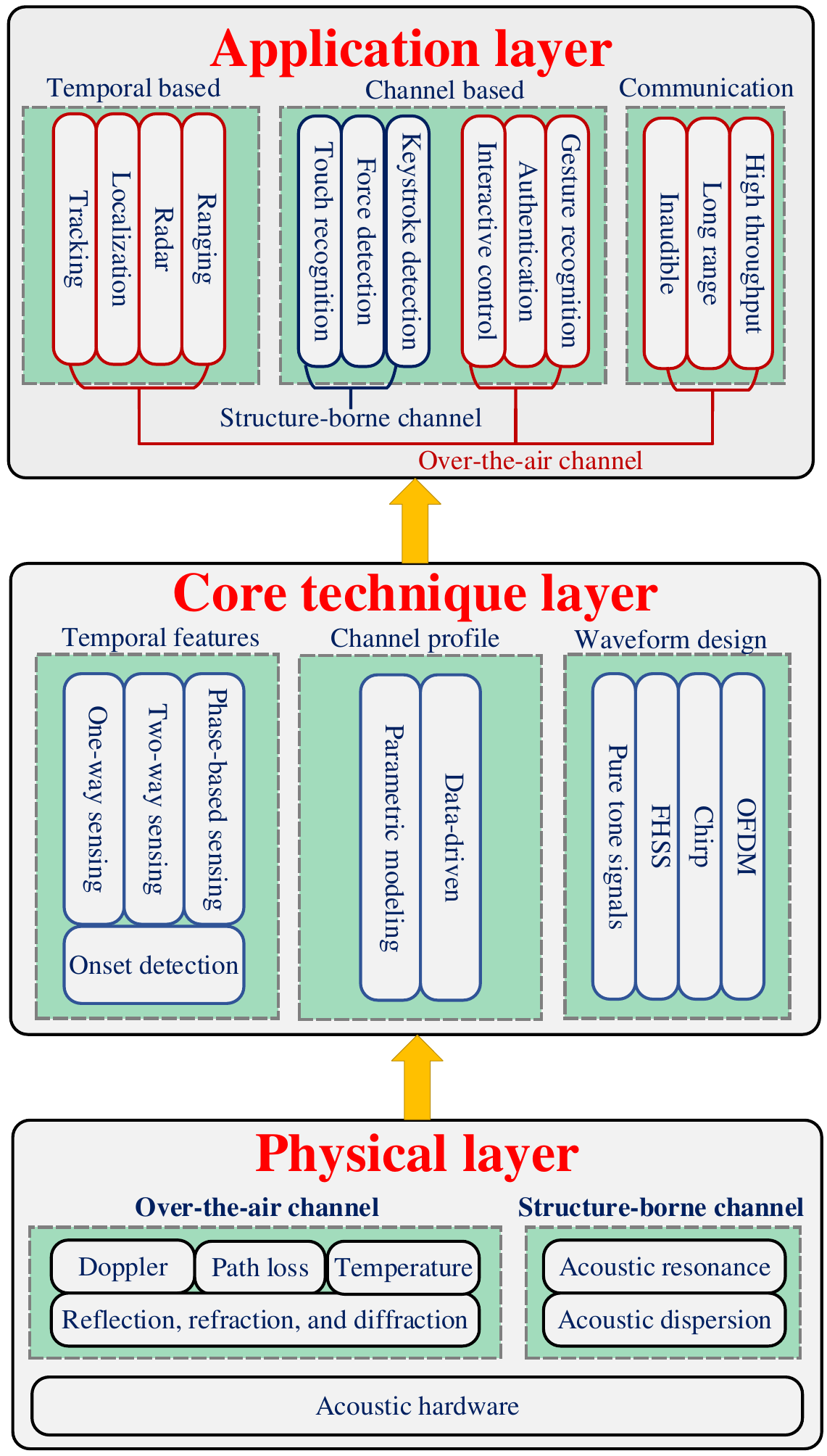}
  \caption{A general framework for acoustic sensing.}
  \label{fig:system overview}
  %\vspace{-2ex}
\end{figure}

Internet of Things (IoT)~\cite{IoTConcept} technologies enable everyday objects to connect and communicate with each other by augmenting them with sensing, processing, and computation units. 
% Such novel architecture sparks a trend to deploy sensing and computation tasks on IoT devices. With the ever increasing computation power and rich built-in sensors, the IoT devices are deployed in more challenging sensing tasks. Such a technology leverages distributed IoT devices to perform challenging sensing and processing tasks.
With the ever increasing computation power and rich built-in sensors available in IoT devices, novel applications emerge by repurposing sensors beyond their primary use. For instance, cameras are intended for taking photos but have been utilized in visible light communication~\cite{VLC}. Gyroscope and accelerometer sensors are designed for attitude estimation but have been used extensively in activity recognition~\cite{Lasagna}. WiFi signals, originally used for communication, have been widely applied in many context-aware computing applications including localization~\cite{WiTrack,Ubicare,Chronous} and gesture recognition~\cite{WholeHomeGesture,CARM}. 
% After years of unremitting efforts by researchers in the community, there are now abundant novel sensing approaches.
In this paper, we target innovative sensing mechanisms that exploit acoustic front-ends on commodity IoT devices. 
% sensors.

Acoustic front-ends, namely microphones and speakers, are among of the most commonly used transducers in IoT devices. They are generally designed for playing back and recording audio signals, but also play a pivotal role in passive sensing applications such as speech recognition~\cite{SpeechRecognition1,SpeechRecognition2} and acoustic source localization~\cite{AudioBeamforming,AudioBeamforming1}.
%In recent years, smart IoT products with acoustic front-ends and cloud-based machine learning technologies are gaining popularity; examples are Google Home~\cite{GoogleHome} and Amazon Echo~\cite{AmazonEcho}. However, these developments are limited to passive acoustic sensing in the human audible frequency range, and thus leave many untapped potentials to be explored.  
%
%by and large, utilize passive sound signals that are recognizable for humans, which are called \emph{passive sensing}. They are encouraging but still exploit acoustic sensors in a traditional way. In addition, they only use a small amount of available acoustic bandwidth. Therefore, we believe there are still many untapped potentials in acoustic sensing.
%
%Consequently, the scope of this survey goes beyond passive sensing to encompass active sensing and communication, yet we still adopt the term - \textit{acoustic sensing} - for the sake of brevity.\footnote{Communication can be deemed as a special type of sensing, as the receiver is just a sensor that ``deciphers'' information out of transmitted waveform.}
%
Novel active sensing mechanisms that treat acoustic front-ends as transceivers to emit and capture wireless signals have gained a lot interests in the research community. 
%Thus, many RF analogous properties of acoustic signals have been used extensively.
For instance, acoustic signals have been used to establish aerial acoustic communication channels to transmit a small amount of information~\cite{AcousticCDMA,ChirpCommunication,AcousticNFC,AcousticOFDM}. 
% Like Radio Frequency (RF) signals, 
Also, the reflective property of acoustic signals have enabled the development of acoustic short-range radars for floor map reconstruction~\cite{BatMapper} and gesture recognition~\cite{FingerIO,LLAP,Strata}. 
%Combining RF analogous properties with its unique features, acoustic signals can be exploited in many more fields. Due to
Moreover, the relatively slow propagation speed of acoustic waves (compared to, e.g., Radio Frequency or RF) in common media allows to achieve comparable performance using a relatively low bandwidth than those with RF technologies. Consequently, it is possible to achieve accurate Time-of-Flight estimations that further support many context-aware applications~\cite{BeepBeep,RFBeep,ALPS,ALPSPre,Liu,Guoguo,ARABIS}.
Last but not least, 
%the unique signatures of acoustic emitting sources can be utilized for authentication or activity recognition~\cite{BreathPrint}, whereas 
active acoustic sensing can enable deformity detection and estimation of non-acoustic emitting objects by transmitting purposefully modulated acoustic signals and make inference based on the reflected waveforms captured by microphones~\cite{SoundWave,VSkin}. 

\begin{figure*}[b]
  \centering
    \subfigure[Sound recording system.]
    {
        \label{fig:a5}
        \includegraphics[width=1.3\columnwidth]{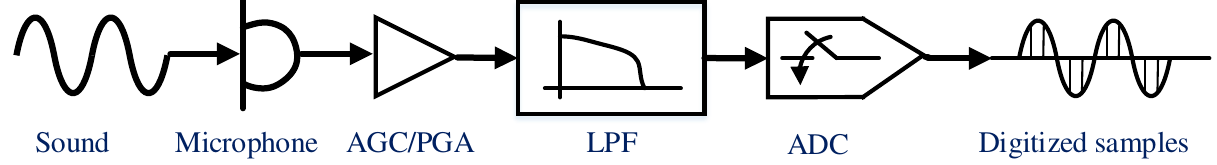}
    }
    \\
    \vspace{0.002\textwidth}
    \subfigure[Sound playback system.]
    {
        \label{fig:b5}
        \includegraphics[width=1.3\columnwidth]{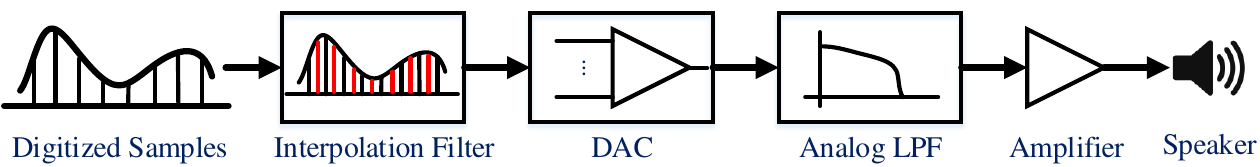}
    }
  \caption{Diagrams for typical acoustic hardware.}
  \label{fig:microphone and speaker}
\end{figure*}

%
% \footnote{It should be noted that communication and sensing are different terminologies. However, in this paper, to fully uncover the potential of acoustic signals, we view acoustic communication as a special type of acoustic sensing, whose purpose is to convey information.}
%
%Passive sensing via unique acoustic signatures for authentication~\cite{BreathPrint} or activity recognition~\cite{DopEnc,BodyBeat} is also feasible. These applications start to utilize active sound (purposefully modulated sound signals) , which is called \emph{active sensing}, and employ inaudible bandwidth. \textcolor[rgb]{1.00,0.00,0.00}{Nevertheless, these novel sensing approaches essentially follow a general paradigm that detects and responds to purposefully modulated acoustic waves or passive sound signals in the environments to extract useful information}\footnote{\textcolor[rgb]{1.00,0.00,0.00}{It should be noted that communication and sensing are different terminologies. However, in this paper, to fully uncover the potential of acoustic signals, we put acoustic communication as subset content of acoustic sensing}.}.

Despite tremendous efforts in developing acoustic sensing applications in the past decade, a systematic treatment of fundamental principles, key design considerations, and innovative methodologies is still missing. As a result, when developing applications based on acoustic sensing, researchers and developers often have to start from scratch and reinvent the wheel. %Therefore, a tutorial or survey that generalizes key building components of acoustic sensing and guide for new research directions is highly desirable.
In this paper, we provide the first systematic survey on recent advances in acoustic sensing with emphasis on novel sensing approaches on commodity hardware (with a bandwidth below ${24}$ kHz), as opposed to those that require special-purposed hardware such as underwater acoustic communication or ultrasonic sensing. We review the relevant research in a bottom-up manner, from the physical layer, core technique layer, and application layer, as shown in Fig.~\ref{fig:system overview}. The physical layer includes basic hardware components, acoustic platforms as well as the air-borne and structure-borne channel characteristics.
% fundamental design considerations including hardware circuit, waveform design, digital modulation, etc; 
The core technique layer encompasses key mechanisms to generate acoustic signals (waveforms) and to extract useful temporal, spatial and spectral information from received signals. 
The application layer builds upon the functions offered by the core techniques to realize different acoustic sensing applications. We group acoustic sensing applications into  aerial acoustic communication, applications leveraging temporal features such as ranging, acoustic radar, acoustic localization and tracking, applications enabled by estimating acoustic channel characteristics such as gesture recognition, speaker liveliness detection and interactive controls. 
% including context-aware application, human computer interaction, and aerial acoustic communication, based on the useful signatures from the processing layer. 
We highlight unique challenges due to the limitations of physical devices and acoustic channels and how they are mitigated or overcame by core processing techniques and application-specific solutions.
Along each category of applications, we discuss research opportunities for further investigation. Finally, we summarize under-investigated areas and emerging applications in acoustic sensing as a whole. 
% and propose a general framework that encompasses main building blocks of typical acoustic sensing systems. This framework provides a novel layered taxonomy of previous work consisting of application layer, processing layer, and physical layer as depicted in . This framework bares similarity with the layered architecture in the ISO/OSI ${7}$-layer reference model~\cite{ISO} for computer networks.
% In the framework, the physical layer focuses on building appropriate setups; converts acoustic signals into digital samples with appropriate hardware and signal processing modules; %It lays the foundations for the sensing system;
%In addition to discussing the technical details, we also highlight limitations in the form of \textit{challenges} for each layer; the (partial) solutions to them appear in upper layers while the unsolved part shall remain as future directions.

The remainder of this paper is organized as follows. In Section~\ref{sec:device or hardware background}, we introduce typical hardware components and system supports offered by commodity devices and the properties of acoustic channels. In Section~\ref{sec:core acoustic mechanisms}, we present the core techniques to generate acoustic signals and to extract temporal and channel features from received signals.
% building blocks for the above applications and discuss key enabling techniques including parametric modeling and data-driven approaches. 
In Section~\ref{sec:solution or application},  a variety of applications are discussed in details according to the core techniques on which they are based;
% existing work according to their application scenarios into three categories, namely, context-aware applications, human-computer interface (HCI), and aerial acoustic communication. 
we discuss research opportunities for each category separately and present future directions in Section~\ref{sec:future direction}.
%we discuss remaining challenges and new research directions in Section~\ref{sec:challenges and future directions}. 
Finally, we conclude the paper in Section~\ref{sec:conclusion}.

%\section{Device/Hardware Background}
\section{Acoustic Devices and Channels}
\label{sec:device or hardware background}
%
%In this section, we set up the foundation for discussing detailed mechanisms and solutions later. In particular, we explain the commonly adopted acoustic hardware along with their influences, and we also present a few popular platforms that support acoustic operations.
%\textcolor{red}{In this section, we present physical layer foundation for the processing and application later. In particular, we explain the commonly adopted acoustic hardware, channel properties, platform diversity, and challenges.}

% Physical layer interfaces with acoustic hardware and the processing layer. It records audio streams and playbacks desired waveform.
% In this section, \textcolor{red}{we introduce background information on acoustic hardware. }
%we discuss physical layer design, in particular, major design issues including hardware issues, receiver design, waveform designs, and bandwidth consideration. 
On an acoustic-enabled device, applications inject processed digital signals to its acoustic frontend (e.g., speakers), which transmits analog acoustic signals through air-borne or structure-borne channels. Upon reception at the acoustic frontend (e.g., microphones) of a receiver device, the analog signals are transformed to digital signals for further processing and are eventually delivered to an application. During the process, extra latency due to processing and system delays, distortions from acoustic frontends and channels and noise and inferences from on-board circuits or environments are introduced. In this section, we present the physical components of acoustic active sensing pipelines and their characteristics. The limitations of acoustic hardware and systems as well as channels pose non-trivial challenges to acoustic signal processing and application development. 

%\subsection{Hardware}
%\label{sec:hardware}
\subsection{Acoustic Hardware}
\label{sec:acoustic hardware}
%
% The typical pipelines of sound recording~\cite{BackDoor} and playback~\cite{AudioHardware} system,
The typical pipelines of \textit{sound recording and playback systems} are shown in
Fig.~\ref{fig:microphone and speaker}(a) and (b) on the transmitter and receiver sides, respectively. The former converts mechanical (acoustic) waves into digital samples while the latter reverses this process. In a recording system, acoustic signals are first converted into voltage signals by a \textit{microphone}. % with a bandwidth normally up to 100~\!kHz~\cite{BackDoor}. 
An Automatic Gain Control (AGC) or Programmable Gain Amplifier (PGA) then amplifies the voltage signals to 
fit the dynamic range of a posterior Analog-to-Digital Converter (ADC); this helps to improve the digitization resolution and to avoid saturation~\cite{Max9814Datasheet}. Amplified signals further go through a Low Pass Filter (LPF), also known as an anti-aliasing filter, and become band-limited signals. The cut-off frequency of the LPF is ${f_s / 2}$, where ${f_s}$ is the sampling rate. Filtered signals are
% go through a buffer and 
finally converted to digital samples by an ADC. A sound playback system reverses the above process. Digital samples are first interpolated and then fed into a Digital-to-Analog Converter (DAC) to become analog signals. The analog signals are further amplified and finally converted to acoustic waves by a \textit{speaker}. 

\begin{figure}[t]
    \centering
    \subfigure[Type A layout.]
    {
        \label{fig:a11}
        \includegraphics[width=0.38\columnwidth]{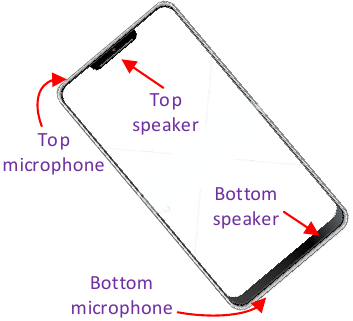}
    }
    \hspace{0.036\textwidth}
    \subfigure[Type B layout.]
    {
        \label{fig:b11}
        \includegraphics[width=0.38\columnwidth]{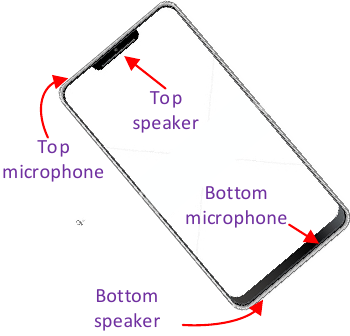}
    }
    \subfigure[Type C layout.]
    {
        \label{fig:a12}
        \includegraphics[width=0.38\columnwidth]{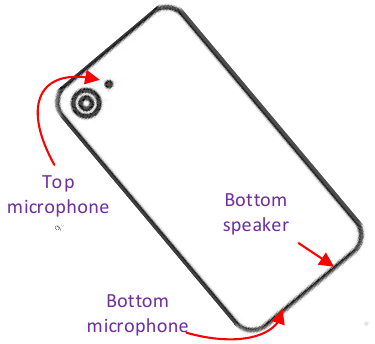}
    }
    \hspace{0.036\textwidth}
    \subfigure[Type D layout.]
    {
        \label{fig:b12}
        \includegraphics[width=0.38\columnwidth]{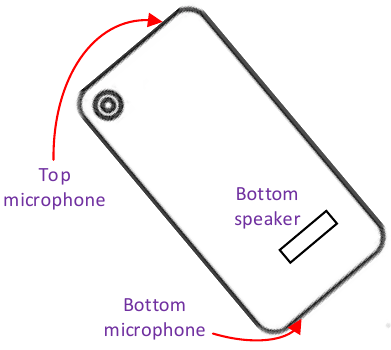}
    }
\caption{Acoustic sensor layouts of different phone models: Front views on (a) and (b) of Type A (e.g., HUAWEI MATE 30 and Honor V20) and Type B (e.g., OPPO A59 and Xiao MI8 SE). Back views on (c) and (d) of Type C (e.g., One Plus 5T) and Type D (e.g., Samsung Galaxy S5 and HUAWEI MATE9).} \label{fig:typical layout}
\end{figure}

Contemporary commodity devices often include more than one microphone and one speaker. For instance, modern smartphones utilize two speakers to play stereo audio and employ two microphones to enhance recording qualities. The typical layouts of microphones and speakers on smartphones are shown in Fig.~\ref{fig:typical layout}. The physical layouts of microphones and speakers are important considerations in some applications, which we defer to Section~\ref{sec:solution or application} for more detailed discussion. 

On acoustic playback systems, amplification or gains of sounds can often be configured. High-power speakers are utilized in applications that operate at long ranges. The most important parameters for acoustic recording systems include sampling rate,  bit resolution, and the number of channels (or microphones). %\textcolor{red}{The sampling rate and the bit resolution can affect the signal quality.} 
The higher the sampling rate and bit resolution, the better the signal quality. The sampling rate typically ranges from 8~\!kHz to 44.1~\!kHz, but certain devices (e.g., high-end smartphones) have a maximum sampling rate of 192~\!kHz~\cite{Pixel}. The most common configuration for bit resolution is 8-bit or 16-bit, but customized devices may support a maximum of 32-bit.
%High bit resolution introduces less quantinization errors and thus preserves more signal details. 
Whereas higher bit resolutions are generally preferred to preserves more signal details, choosing a suitable sampling rate and making use of the two channels commonly available to both sound playback and recording systems have a profound impact on the sensing performance and are further discussed in Section~\ref{sec:future direction for temporal features}. 

The acoustic hardware on commodity devices often introduce non-flat frequency response on its input signals. The non-flat frequency response, also known as \textit{frequency selectivity}, describes a phenomenon where acoustic signals experience different channel gains at different frequencies.  This can be problematic for signals that occupy a relative wide frequency range.
On the receive side, signals whose bandwidth is below 8~\!kHz~\cite{AcousticNFC,BatMapper} can have a higher receiver gain while those above experience sharp attenuation since microphones on commodity devices are designed primarily for recording human voices.  
On the transmitter side, with \textit{speaker diaphragm inertia}~\cite{Inertial}, introduced by a non-electronic component (diaphragm) in speakers, the movement of speaker diaphragms cannot catch up with fast changes in input signals. Thus, higher frequency signals tend to be dampened. 
Additionally, it causes ringing effects~\cite{ChirpCommunication} or frequency  leakage~\cite{ForcePhone}. Ringing effects describe the problem in the time domain where a transmission has a delayed start and/or a prolonged transmission duration. In contrast, frequency leakage describes the problem in frequency domain where a transmission of a band-limited signal can cause out-of-band artifacts. As a result, speaker diaphragm inertia can generate audible noise even when the transmitted signal only occupies inaudible frequency bands.\footnote{Though the audio range falls between 20~\!Hz and 20~\!kHz, most people normally do not hear sounds about 18~\!kHz~\cite{AcuTe,VSkin,CAT}. Therefore, the frequency range between 18 and 20~\!kHz are often deemed \textit{inaudible} and hence commonly used for acoustic communication and sensing.} 

\subsection{Acoustic Channel Properties}
\label{sec:acoustic channel property}

%\textcolor{red}{We talk about properties of acoustic channel that includes acoustic front-ends and the medium acoustic signals propagate through. 
As acoustic signals generated by mechanical vibrations can propagate through both air, liquid, and solid medium, we next discuss the properties of two channels, i.e., air and solid medium that are relevant to acoustic sensing on commodity devices. The acoustic signals propagate through air and solid medium are called \textit{Air-borne Signals} (AS) and \textit{Structure-borne Signals} (SS), respectively. 

\subsubsection{Characteristics of Air Channels} When acoustic signals propagate through air, they radiate their energy spherically towards surroundings and can be deemed as ``wireless signals''. Hence, they share many similar features as radio signals. Specifically, acoustic signals experience path loss, Doppler effects, reflection, refraction, and Diffraction. 

\begin{itemize}
    \item \textsl{Path Loss}:  Path loss describes the phenomenon that the energy of acoustic signals decrease over the propagation distance. Specifically, the acoustic intensity, measured as the ratio between power and area, is inversely proportional to the square of  distance~\cite{Pathloss}. Supposing the intensity of a particular acoustic signal is denoted by $p$, then {a free-space path loss model is given by}~\cite{Pathloss}:
    \begin{equation}\label{eq:path loss}
        p = k\frac{p_0}{r^2},
    \end{equation}
    where $k$ is a coefficient, $p_0$ is the intensity when distance to the source is $r = 0$.
    \item \textsl{Doppler Effects}: 
Doppler Effect refers to the change in wave frequency during the relative motion between an acoustic source and its receiver (or observer). In particular, the Doppler shift in frequency is given by~\cite{DopplerEffect}:
    \begin{equation}\label{eq:doppler effect}
        \Delta f = \frac{\Delta v_d}{c}f,
    \end{equation}
   {where $f$ is the original frequency, $\Delta v_d$ denotes the relative speed between the transmitter and the receiver, and $c$ is the speed of sound in air.} 
    \item \textsl{Temperature Effect}: Unlikely RF waves, the propagation speed of acoustic signals is temperature dependent given by ~\cite{SonicThemometer}:
    \begin{equation}\label{eq:speed temperature relationship}
        c = \sqrt{403T_{p}\left( {1 + 0.32e/\rho} \right)},
    \end{equation}
    where $c$ is the sound speed (m/s) in air, $T_p$ is the temperature (in Kelvin), $e$ is the vapor pressure of water in air, $\rho$ is the absolute atmospheric pressure. From Eqn.~\eqref{eq:speed temperature relationship}, we see that sound speed is also dependent on humidity. However, since $e/\rho$ tends to be quite small, Eqn.~\eqref{eq:speed temperature relationship} can be further simplified as ${c^2} = 403T_{p}$. 
    \item \textsl{Reflection, Refraction, and Diffraction}: Reflection means that when an acoustic signal hits a boundary of two different mediums, it will be reflected back. The amount of reflected signals depends on the level of difference between the two mediums. Reflections from solid surfaces in the environments contribute to multipath effects. Refraction happens when a fraction of acoustic signals propagate into another medium and change their direction. Diffraction involves a change in direction of waves as they pass through an opening or around a barrier in their path. This happens when obstacles are smaller than the wavelength of the wave.
\end{itemize} 

\subsubsection{Solid Medium Property} {Acoustic signals in solid medium exhibit much differences with those in air. Next, we highlight two key properties within solid medium,} namely, acoustic dispersion and resonance. 

\begin{itemize}
    \item \textsl{Acoustic Dispersion}: As an acoustic signal containing a rich set of frequency components propagates in a solid medium, high frequency components travel faster than low frequency ones. The speed $c_f$ of a specific frequency component $f$ is given by~\cite{AcousticPhysics}:
    \begin{equation}\label{eq:acoustic dispersion equation}
        {c_f} = \sqrt[4]{{\frac{{Eh{f^2}}}{{12\rho \left( {1 - v_p^2} \right)}}}},
    \end{equation}
    where $v_p$ is the phase velocity, and $E, \rho, h$ are constants that characterize a medium: $E$ quantifies elasticity, $\rho$ characterizes stiffness, and $h$ represents thickness. Fig.~\ref{fig:acoustic dispersion} shows the signal recorded by an earbud when a person taps her teeth and generates a pulse vibration that travels through her jaw and skull bones. 
    \begin{figure}[h]
    \centering
    \includegraphics[width=3.5in]{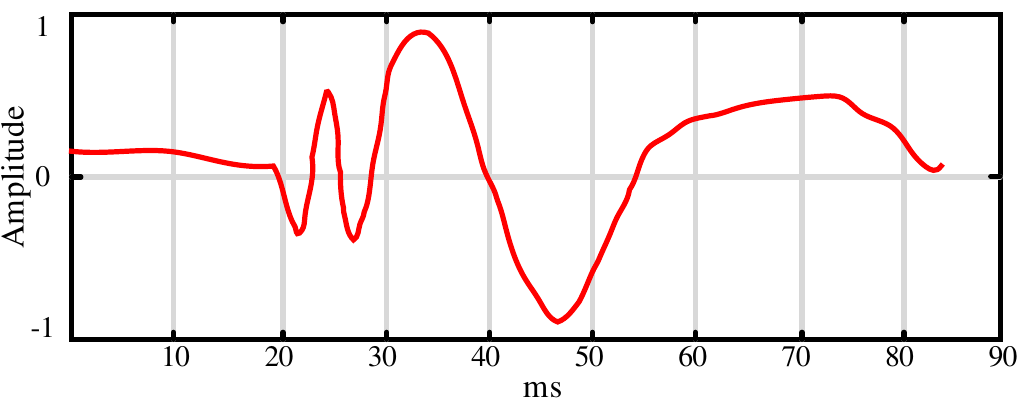}\\
    \caption{Acoustic dispersion~\cite{EarSense}. The waveform is recorded by an earbud (with on-board microphone) when a user taps a tooth.}
    \label{fig:acoustic dispersion}
    \end{figure}
    \item \textsl{Acoustic Resonance}: Acoustic resonance describe the phenomenon that a solid medium selectively amplifies a particular frequency component while attenuating other ones. This particular frequency is called resonant frequency, which changes  if the properties or the state of the solid medium differs, say, under external forces or undergoing shape transformations. 
\end{itemize}
Besides the afore-mentioned difference, acoustic wave propagation in air and solid medium also differ in speed. The speed of AS is 340~\!m/s (at temperature of 25$^\circ$) and for SS, the speed can be over 2000~\!m/s (depending on the type of medium)~\cite{VSkin}. 
Interestingly, an acoustic event can generate both AS and SS at the same time. SS appears ahead of the AS in the captured signals if a receiver is in contact with the surface that vibrates and generates the acoustic event. However, depending on the degree of coupling, with the same sensing device, the signal intensity of SS tends to be significantly lower than AS. 

%AS and SS also have some similarities or connections. 
%For instance, both AS and SS can be utilized as  \textit{acoustic fingerprint} to represent a particular state of a channel or a special channel. 
%As another example, an acoustic event would generate both AS and SS at the same time. SS appears ahead of the AS in the captured signals if the detection sensor is in contact with the surface that vibrates and generates the acoustic event. More discussions on the complementary utilization of both AS and SS can be found in Section~\ref{sec:future direction for temporal features}.

%\subsection{Hardware Influence}
%\label{sec:hardware influence}
%

% and more discussions on this can be found in \textcolor{green}{Section~\ref{}.}

\subsection{Acoustic Platforms}
\label{sec:platform diversity}
Due to the pervasiveness of acoustic transducers and IoT devices, platforms that support acoustic operations are diverse, ranging from customized embedded devices and wearables to general-purpose smartphones and laptops. On general-purposed devices, which reply on commodity operating systems (OS) such as Windows, Linux, Android, MacOS and iOS, acoustic applications are subject to limitations of available application programming interfaces (API) and  need to contend with other applications or system services in resources.

%To capture and playback arbitrary signals on a general-purpose OS, one can either use existing software utilities such as \textsf{Audacity}~\cite{Audacity} for offline processing or deal with raw samples using platform-dependent APIs~\cite{AudioTrack}. 
%Nevertheless, one should be very careful in choosing appropriate software utilities and APIs to retrieve clean raw samples, because they may potentially introduce internal ``distortions''~\cite{LibAS,UbiTap}. %For instance, on Android platforms, the class named AudioRecord is in charge of audio recording and it has a enum to configure its input sources. If the input source is configured as AUDIO.UNPROCESSED, the captured signals would suffer from no internal ``distortions'' that may vary from device to device. 
%One thing worth noting is that acoustic hardware often has uncertainty system delay (typical 10~\!ms) in response to software signaling. This can be notably detrimental to time sensitive \textcolor{green}{applications presented in Section~\ref{fig:time sensitive application}}. Deploying applications at kernel layer to sidestep this system delay is feasible but this may require platform-dependent kernel hacking, incurring another platform diversity problem. Designing algorithms that are agnostic to system delay may be other potential approaches.

\begin{table*}[!b]
\centering
\caption{Comparison of different waveforms designs for sensing purpose.}
\label{tab:waveform design comparision}
\begin{tabular}{|c|c|c|c|}
\hline
\textbf{Waveform} & \textbf{Advantages}  & \textbf{Disadvantages} & \textbf{\makecell{Suitable applications}}\\ \hline
Pure tone & \makecell{Doppler-aware, Responsive,\\less computational requirements} & \makecell{Vulnerable to interference} & Gesture recognition/tracking (relatively short range) \\ \hline
FHSS modulated & \makecell{less correlation sidelobes,\\good multipath resolution} & \makecell{Sensitive to noise} & Gesture tracking (relatively short range)\\ \hline
Chirp & \makecell{Noise-resilient, multipath resilient} & \makecell{Computation intensive} & Ranging, localization, radar (relatively long range)\\ \hline
\end{tabular}
\end{table*}

Different platforms have their advantages and disadvantages in developing acoustic sensing solutions. Platforms running Windows and Linux offer a wide range of software utilities that allow developers to focus on algorithm design and hence facilitate fast prototyping~\cite{LibAS}. To capture and playback acoustic signals, one can either use existing software utilities such as \textsf{Audacity}~\cite{Audacity} for offline processing or directly deal with raw samples using platform-dependent APIs~\cite{AudioTrack}.  However, systems running these OSs tend to be less portable. In contrast, mobile and wearable devices powered by Android or iOS platforms offer native acoustic APIs and are convenient for carrying out various experiments in both indoor and outdoor settings. General-purpose OSs can 
introduce uncontrollable system delays (in tens of millisecond), missing acoustic samples (e.g., due to insufficient buffers) and  ``distortions'', say dynamic gains hence inconsistent acoustic fingerprints, as the result of automatic gain control~\cite{LibAS,UbiTap}. 
%XXX: examples of distortions? 
These artifacts should be taken in account or compensated in developing acoustic solutions. 
Lastly, customized embedded platforms using commodity hardware components generally allow flexible physical layer configuration, including increasing the number of channels and tuning hardware properties~\cite{SST}. 

Platform diversity implies that it difficult if not impossible to have one-size-fit-all solutions. When porting  applications from one platform to another~\cite{ForcePhone}, one also needs to devise calibration procedures to mitigate such diversity. 

%On any platform, the acoustic hardware has an inherent latency, known as \textit{system delay}, in response to software signaling. \textcolor{red}{For instance, sound may be produced after a short delay once a play command is issued in the application layer.}
%Typical, the system delay on popular Android platform is around 10~\!ms~\cite{AndroidAudioProblem} depending on system load~\cite{BeepBeep}. This uncertain delay is caused by operating system (OS) being non-responsive when waiting for the right scheduling cycle to perform I/O operations; it then leads to inevitable latency in responding application commands. Although one can hardly notice this delay, it may significantly damage timing measurements and is hence notable detrimental to time-sensitive applications.

\subsection{Challenges}
\label{sec:challenges on devices}
The characteristics of acoustic hardware, channels and platforms are key design considerations in acoustic sensing solutions. Combating the imperfection and limitations of physical layers requires addressing the following challenges:  

%Given the compact and low-cost nature of the commodity devices, the acoustic hardware offered by them inevitably raises some fundamental challenges listed as follows.
%
\begin{itemize}

    \item Frequency selectivity  caused by acoustic transducers and channels unavoidably results in signal distortions that affect system performance.
   
   \begin{challenge}
   \label{clg:freq_sel}
       Handling frequency selectivity induced signal distortions.    
   \end{challenge}
    %Solutions to this problem may require
    Various compensation approaches can be applied to flatten the frequency response.
    % the frequency selectivity may be feasible solutions
    More details on this are presented in Section~\ref{sec:cir representation}.
    \item 
    Speaker diaphragm inertia mentioned in  Section~\ref{sec:acoustic channel property} may cause disruptive audible noise even if the modulated signal is in inaudible frequency ranges. 
\begin{challenge}
\label{clg:audibility}
    Suppressing audible noise incurred by acoustic transmissions.
\end{challenge}

Maintaining smooth amplitude and phase transitions between each sample is crucial to mitigate this problem and more relevant techniques will be elaborated in Section~\ref{sec:tracking} and Section~\ref{sec:future direction for channel characteristics}.    
\item As explained in Section~\ref{sec:platform diversity}, the uncertain system delay in response to software signaling can be detrimental to the extraction of temporal characteristics,
    % This uncertainty latency, known as system delay, can be notably detrimental to time sensitive applications presented in \textcolor{green}{Section~\ref{fig:time sensitive application}}. %due to its complex hierarchical software architecture. 

%    \begin{displayquote}
        \begin{challenge}
        \label{clg:system_delay}
            Masking the uncertainty system delay for the upper layer functionalities.
        \end{challenge}
%    \end{displayquote}  
    % Possible solutions include deploying application at kernel layer to sidestep this system delay but this may require platform-dependent kernel hacking, incurring another platform diversity problem. 
    Designing algorithms that are agnostic to system delay may be potential approaches and more discussions on this can be found in Section~\ref{sec:basic timing measurements} and~\ref{sec:phase-enabled accurate timing}.
\item Though most acoustic devices share the similar hardware pipeline (Fig.~\ref{fig:microphone and speaker}), the diversity in both acoustic front-ends and software platforms can pose challenges to application deployment. 

\begin{challenge}
\label{clg:calibration}
Reducing or eliminating laborious device-dependent calibration when developing acoustic applications.
\end{challenge}

Tackling this challenge may require certain level of platform homogenization, and more discussions  can be found in Section~\ref{sec:channel characteristics-enabled applications}.
    % [CC]: platform related content is currently not provided.
%
\end{itemize}

\begin{table*}[!b]
\centering
\caption{{Comparison of different waveforms for communication.}}
\label{tab:waveform design comparision for communication}
\begin{tabular}{|c|c|c|c|}
\hline
\textbf{Waveform} & \textbf{Advantages}  & \textbf{Disadvantages} & \textbf{\makecell{Suitable applications}}\\ \hline
\makecell{Pure tone,\\FHSS modulated} & \makecell{High data rate} & \makecell{Short communication range,\\high BER} & Near field communication\\ \hline
%FHSS modulated & \makecell{less correlation sidelobes,\\good multipath resolution} & \makecell{Sensitive to noises} & Gesture tracking (relatively short range)\\ \hline
Chirp & \makecell{Noise-resilient, multipath resilient,\\long communication range,\\low BER} & \makecell{Low data rate} & Long-range, low-volume, strong interference\\ \hline
\end{tabular}
\end{table*}

\section{Core Acoustic Sensing Techniques}
\label{sec:core acoustic mechanisms}
In this section, we present core techniques for acoustic sensing. These techniques mitigate some of the challenges discussed earlier and serve as building blocks for various application.  Waveform design for sensing and communication is first introduced, followed by mechanisms extracting temporal and frequency domain characteristics.

\subsection{Waveform Design}
\label{sec:waveform design}
Waveform design is critical for acoustic sensing systems. 
We draw a distinction between sensing and communication as the designs suitable for them (albeit related) can be rather different. 
The available bandwidth for transmitted acoustic waveforms  normally goes up to 24~\!kHz. The frequency range from $20$~\!Hz to 18~\!kHz is \textit{audible} to human but has better channel gains in commodity acoustic front-ends~\cite{VSkin}, while the \textit{inaudible} range from 18 to 24~\!kHz, is preferred due to less interference but is subject to sharp attenuation.

\subsubsection{Waveforms for Active Sensing}
\label{sec:waveform for sensing}
The most commonly used signals for sensing include \textit{Pure Tone}, \textit{Frequency Hopping Spectrum Spread} (FHSS), and \textit{Chirp}. Their respective advantages, disadvantages, and suitable applications have been summarized in TABLE~\ref{tab:waveform design comparision}. 
\begin{figure}[!t]
    %\vspace{-2ex}
    \centering
    \subfigure[Pure tone signal.]
    {
        \label{fig:pure tone waveform}
        \includegraphics[width=0.44\columnwidth]{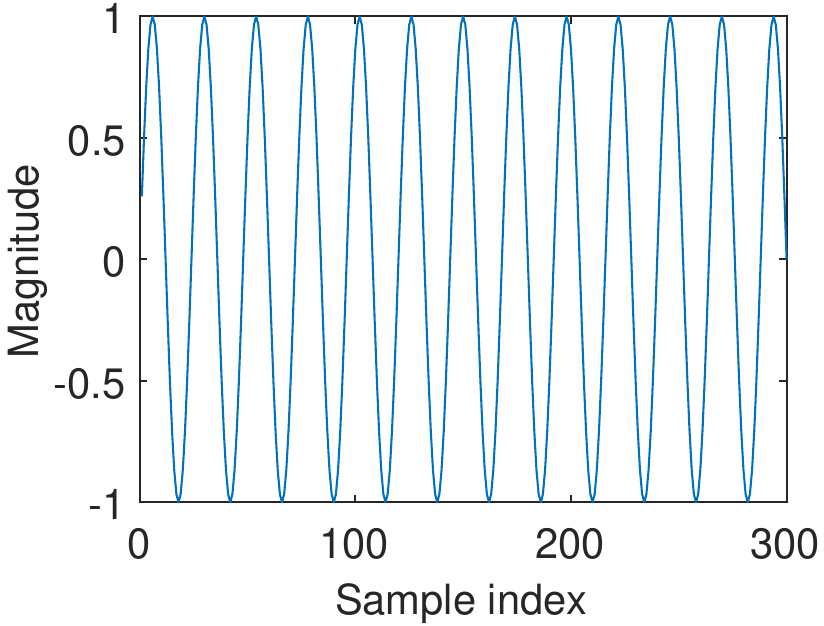}
    }
    \hspace{0.002\textwidth}
    \subfigure[Spectrogram of pure tone signal.]
    {
        \label{fig:pure tone stft}
        \includegraphics[width=0.45\columnwidth]{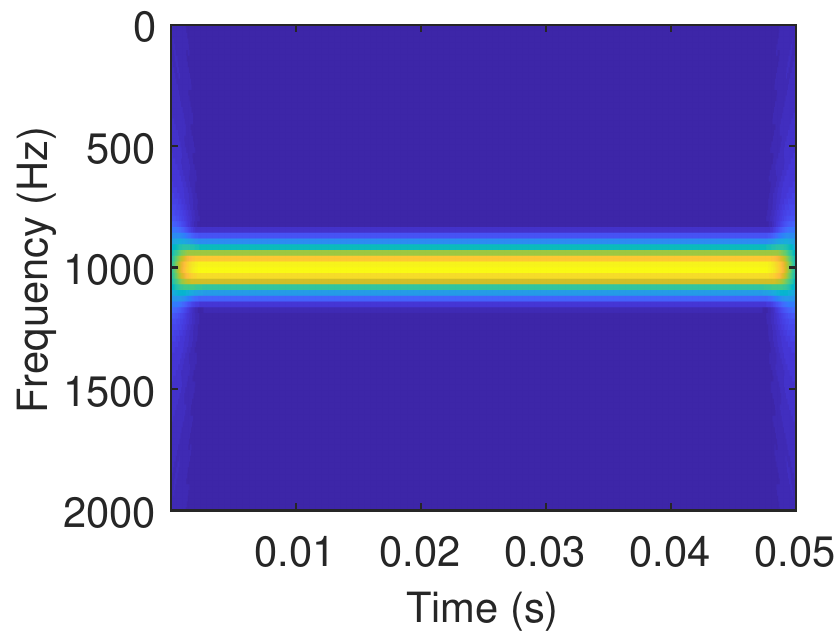}
    }
    \subfigure[Baseband ZC signal.]
    {
        \label{fig:baseband ZC}
        \includegraphics[width=0.44\columnwidth]{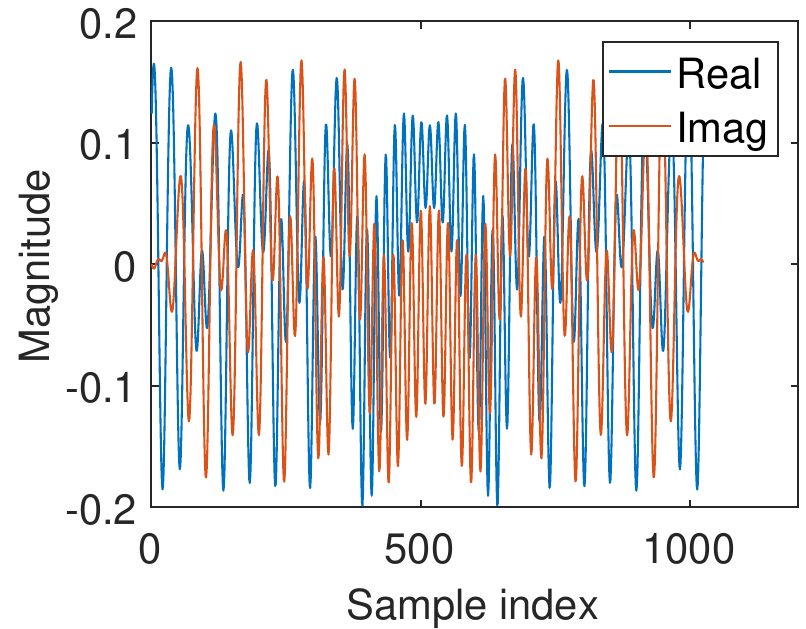}
    }
    \hspace{0.002\textwidth}
    \subfigure[Spectrogram of ZC signal.]
    {
        \label{fig:ZC stft}
        \includegraphics[width=0.45\columnwidth]{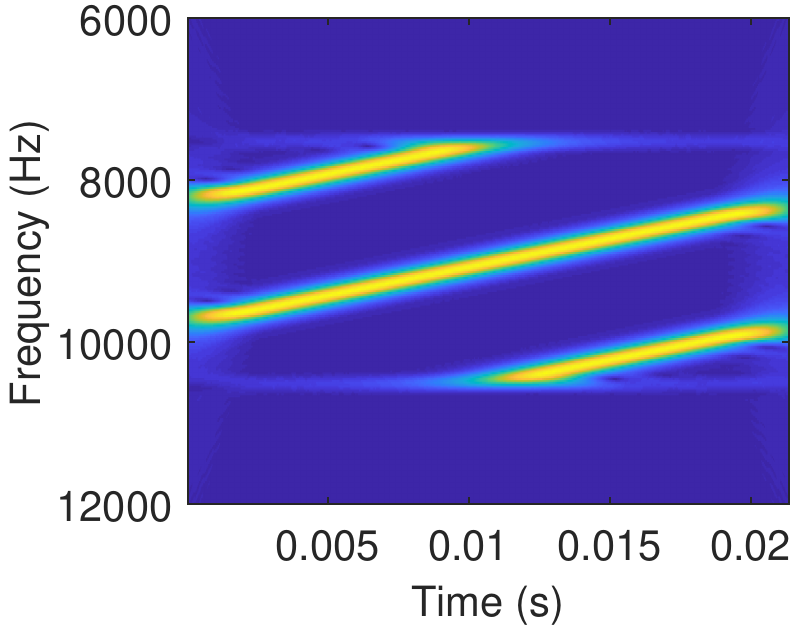}
    }
    \subfigure[Time domain chirp signal.]
    {
        \label{fig:chirp waveform}
        \includegraphics[width=0.44\columnwidth]{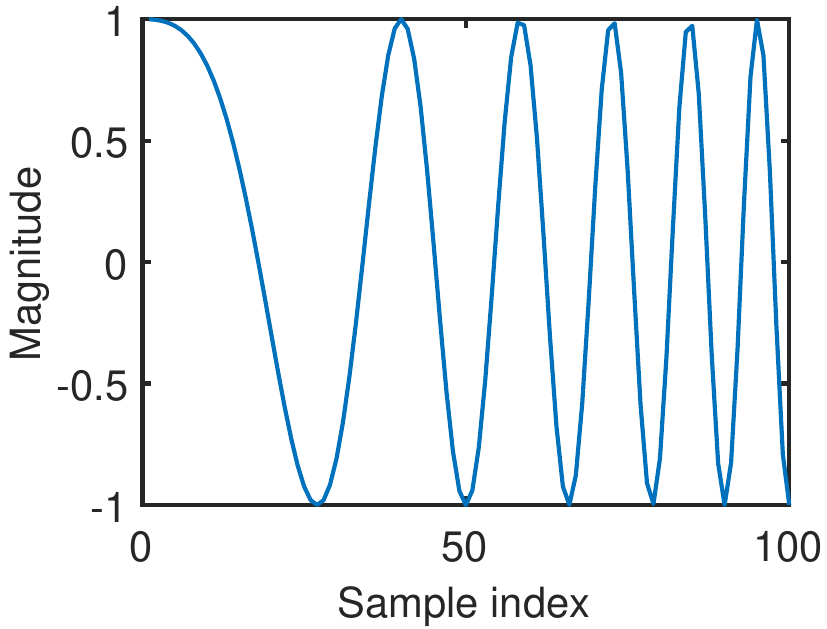}
    }
    \hspace{0.002\textwidth}
    \subfigure[Spectrogram of chirp signal.]
    {
        \label{fig:chirp stft}
        \includegraphics[width=0.45\columnwidth]{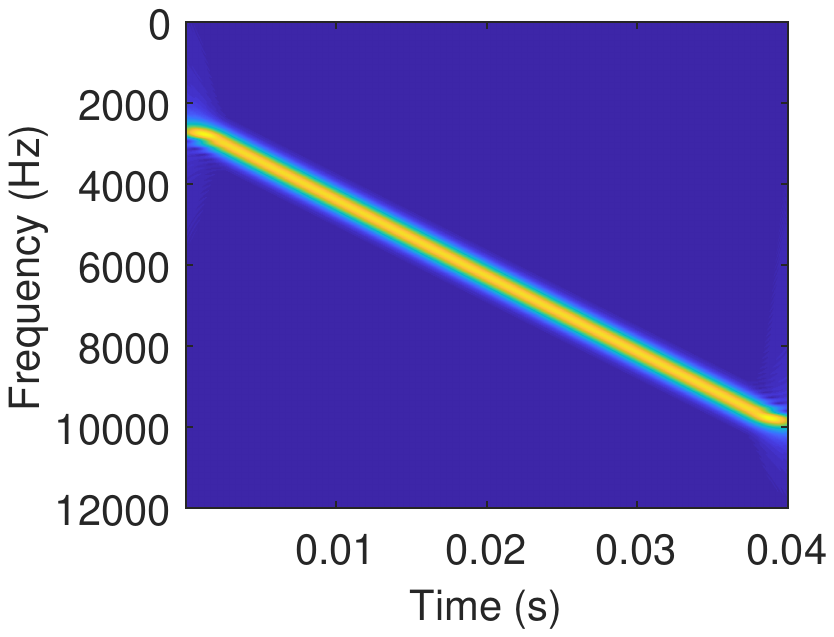}
    }
\caption{Time and frequency domain signal representations. }
\label{fig:signal representation}
\end{figure}

\paragraph{Pure Tone Signals} Pure tone signals are commonly used in acoustic sensing due to their low complexity and high resolutions in tracking Doppler shifts. Their  waveform is represented as ${s\left(t\right) = \cos\left( 2\pi ft + \phi \right)}$ where ${f}$ and ${\phi}$ are the frequency and the initial phase. The waveform and its Short Time Fourier Transform (STFT) spectrogram of a single tone signal are shown in Fig.~\ref{fig:pure tone waveform} and \ref{fig:pure tone stft}, respectively. 
Consider that a moving target transmits a pure tone signal with frequency ${f}$, and the detected Doppler frequency is ${f_\mathrm{shifted}}$ at the receiver side. The relative moving speed can thus be estimated as ${v = \frac{f_\mathrm{shifted} - f}{f}c}$, where ${c}$ is the sound speed. Due to the low propagation speed of sound, it is feasible to achieve a cm/s-level estimation accuracy.  
Moreover, 
% these signals allow for fast and precise phase estimation. In particular,
if phase components across multiple pure tones are available, the phase diversity yields multiple constraints to obtain more accurate phase and frequency estimations~\cite{CRT,CRT1}. 
% facilitating range or localization applications~\cite{Chronous,LLAP}. 
%Finally, the frequency component of a pure tone signal can be easily detected and accurately estimated, enabling tone-based modulation. O
As phase and frequency shifts are correlated with spatial quantities such as range and speed, pure tone signals have been extensively used in localization and tracking applications ~\cite{Chronous,LLAP,AAMouse,CAT}, as well as gesture recognition~\cite{SoundWave,LLAP}. However, they are not suitable for extracting precise timing information due to the periodicity and shallow peaks of auto-correlation functions. 
\paragraph{FHSS Signals} 
%As the low-complexity pure tone signals do not have a satisfactory auto-correlation property, they typically end up \textcolor{red}{with an inability to resolve multipath effect.} 
FHSS modulation rapidly changes the carrier frequency among many distinct frequencies occupying a large spectral band. The changes are controlled by a code or a spreading sequence. 
%To this end, one often resorts another commonly used waveform generated via FHSS.
% modulation. FHSS can expose even better spatial quantities than pure tone signals. 
% whose orthogonal sequence is modulated through Binary Phase Shift Keying (BPSK). 
The key design consideration of FHSS modulation is to choose an appropriate orthogonal sequence. Commonly used ones for acoustic applications include Zadoff-Chu (ZC)~\cite{LTE,ZC1,ZC2,ZC3}, Barker code~\cite{VSkin}, GSM training sequence~\cite{Strata}, and m-sequence~\cite{SwordFight}. A baseband ZC signal and its STFT spectrogram after modulation are shown in Fig.~\ref{fig:baseband ZC} and~\ref{fig:ZC stft}, respectively. 
Thanks to the use of spectrum spreading sequences, correlating FHSS modulated signals results in sharper peaks and smaller sidelobes compared to pure tone or even chirp signals~\cite{VSkin}, % thus enabling finer multipath resolution, an edge to develop high-accuracy tracking applications~\cite{FingerIO}. 
%A comparison of auto-correlation between ZC modulated signals and chirp signals is shown in Fig.~\ref{fig:auto correlation comparison}. It is evident that ZC outperforms chirp signals for its more sharp correlation peak. 
making {multiple acoustic reflections readily distinguishable}. Therefore, FHSS modulated signals are a popular choice for high-accuracy gesture tracking~\cite{VSkin,Strata}. Unfortunately, due to their high sensitivity, they can be easily corrupted by channel noises including path loss, the Doppler effect, background interference, etc, making them only suitable for short-range sensing. 
%Unfortunately, they are vulnerable to background noise and are sensitive to Doppler effects~\cite{ChirpCommunication}.
%, making it necessary to look for more robust waveform \textcolor{red}{in the face of background interference}.  
%XXX: still have doubts about noise sensitivity
%[CC]: have changed the statement
%\begin{figure}[t]
%  \centering
%  \includegraphics[width=3.0in]{Figs/correlation_comp.pdf}
%  \caption{Correlation results comparison.}
%  \label{fig:auto correlation comparison}
%\end{figure}

% [CC]：写信号的逻辑是先理论后应用，这一原则在后面的写法中要保持, 在写processing层时保持这一结构。

\paragraph{Chirp Signals} 
% The final most commonly use signals are chirp signals. 
A chirp is a signal in which the frequency increases (up-chirp) or decreases (down-chirp) with time.
A linear chirp 
% also known as Linear Frequency Modulated (LFM) signal, 
is represented by ${s\left( t \right) = A\cos \left( {2\pi \left( {{f_{\min }}t + \frac{k}{2}{t^2}} \right) + \phi } \right)}$, where ${f_\mathrm{min}}$ is the initial frequency, ${A}$ is the maximum amplitude, ${\phi}$ is the initial phase, and  $k=\frac{B}{T}$ is the \textit{modulation coefficient} or \textit{chirp rate}~\cite{AcousticCDMA,BeepBeep,AudioGest,BatMapper}. In sensing, chirp signals are transmitted repeatedly, and thus they are also known as Frequency Modulated Continuous Wave (FMCW). The time and frequency domain representations of a chirp signal are shown in Fig.~\ref{fig:chirp waveform} and~\ref{fig:chirp stft}, respectively.
Auto-correlating chirp signals results in sharp and narrow peaks with 3dB temporal bandwidth inverse proportional to the signal bandwidth, a property also known as {\it pulse compression}. Since the energy of a signal does not change during pulse compression, concentration of its signal power within a narrow interval leads to a peak SNR gain proportional to the product of signal bandwidth (B) and duration (T)~\cite{ALPS}.
As a result, acoustic sensing systems employing chirp signals enjoy three major advantages.
First, chirp signals are robust to dynamic channel conditions such as Doppler effects.
Second, they are detectable even when the signal power is under noise floor~\cite{AcousticCDMA}, and thus can combat strong background noise or  interference.
Third, chirp signals are resilient to multipath fading, allowing extraction of multipath components~\cite{FollowMeDrone}. 
% It is feasible to distinguish multiple reflections with appropriate signal processing models~\cite{FollowMeDrone}. 
%Chirp signals can also be pieced together, forming another widely used signals called Frequency Modulated Continuous Wave (FMCW). 
Thanks to these desirable features, chirp signals have been widely used as a building block in temporal feature extraction~\cite{CAT,FollowMeDrone} and channel characteristics estimation~\cite{TouchActive,ForcePhone}, as discussed in more details in Section~\ref{sec:temporal features} and \ref{sec:channel characteristics}.
% in parametric modeling and act as probe signals in data-driven approaches.

%\paragraph{Challenges on Waveform Design} 
%\textcolor{green}{The aforementioned three popular waveforms have their own advantages in acoustic sensing as we outlined in TABLE~\ref{tab:waveform design comparision}. Generally speaking, pure tone incurs \textit{low system complexity}, FHSS gain \textit{high sensitivity}, and chirps are more \textit{robust}. This in turn reveals that when design a waveform for acoustic sensing, we have to trade off between the above advantages. Consequently, though desirable, 
%\begin{displayquote}
%        \textit{\textbf{Challenge-V}: it is significantly challenging to design a one-fit-all solution that has all the aforementioned advantages.}
%\end{displayquote}    
% Experience gained from RF design may be helpful but needs more investigations. }

%It should be noted that when there needs to transmit modulated signals continuously, it it better to insert a guard interval between adjacent signals so as to mitigate inter-symbol-interference. The underlying fact is that acoustic reflections have severe distortions after a short duration, say $25$~\!ms~\cite{ForcePhone}. As a result, inserting a guard interval could reduce the impact from previous delayed signal block on its subsequent one.

\subsubsection{Waveforms for Communication}
\label{sec:waveform for communication}
The aforementioned waveforms are applicable to aerial acoustic communications as well but different signal configurations (e.g., frequency, spread sequence or chirp rate) represent different source symbols. A comparison of these waveforms and their suitable applications in communication can be found in TABLE~\ref{tab:waveform design comparision for communication}. Commonly used modulations in acoustic communication include Frequency Shift Keying (FSK), Orthogonal Frequency Division Multiplexing (OFDM), and Chirp Spread Spectrum (CSS).

\paragraph{FSK}
As the amplitude and phases of acoustic signals are prone to channel distortions, modulation techniques based on these two features (e.g., Amplitude Shift Keying or Phase Shifted Keying) are not reliable~\cite{ChirpCommunication}. In contrast, 
FSK, a modulation technique that uses pure tone signals with distinct frequencies to represent data bits, is more robust to channel distortion and interference. It can be demodulated via FFT analysis, Hilbert Transform, or coherent detection. Though FSK has low complexity in both modulation and demodulation design, its achievable data rates are low~\cite{MarrayFSK}.

\paragraph{OFDM}
Compared to single-carrier FSK, OFDM is theoretically  more efficient by modulating data symbols on multiple orthogonal subcarriers. It has the potential to achieve a higher throughput under the same bandwidth~\cite{OFDM,OFDMStructure}. However, when implemented in software for acoustic communication, its advantage cannot be fully realized. Specifically, advanced processing modules such as carrier sensing and carrier frequency offset correction, which are common in RF-based OFDM systems, require hardware level modification due to tight timing requirements. Moreover, as OFDM only maps bit streams onto individual subcarriers, a separate modulation scheme is needed to encode these bit streams. High-order modulations such as Quadrature Amplitude Modulation (QAM) are infeasible in acoustic communication due to significant channel distortions.
% Quadrature Amplitude Modulation 
%regardless of what signal features are used, are rather unlikely to be applicable. 
As a result, 
OFDM-based systems achieve similar throughputs as those based on FSK. Both FSK and OFDM are only suitable for short-range communication between stationary devices~\cite{AcousticNFC}.
% This renders acoustic OFDM achieve less comparable throughput than its RF counterparts.

%A typical function block for acoustic OFDM is shown in Fig.~\ref{fig:acoustic ofdm} \cite{OFDMStructure} and the corresponding data flow works as follows. 
%The typical signal processing steps for acoustic OFDM work as follows. First, bit streams are processed by channel coding techniques such as forward error correction and cyclic redundancy check. This step adds redundant information to the original data streams and makes them more noise resilient. Afterwards, the bit streams are parallelized and go through Inverse Fast Fourier Transform (IFFT). This operation generates ready-to-transmit time domain signals. To mitigate inter-symbol-interference (ISI) and inter-channel interference (ICI), cyclic prefix/suffix (CP/CS), a fractional repetition of signal block, is inserted. At this point, an OFDM packet is generated. To make the OFDM packet easy to be detected, a preamble (usually a chirp signal) is inserted before the packet. Finally, the signals are transmitted through the acoustic channels. A receiver reverses the above process. OFDM though can achieve higher throughput than FSK but is only suitable for short-range communication~\cite{AcousticNFC} and suffer from degraded performance in mobile scenarios. 
%Both FSK and OFDM are only suitable for short-range communication~\cite{AcousticNFC} and suffer from degraded performance in mobile scenarios. In comparison, CSS~\cite{CSS} is more reliable and can be used in long-range communication.

\begin{figure}
  \centering
  \includegraphics[width=3.4in]{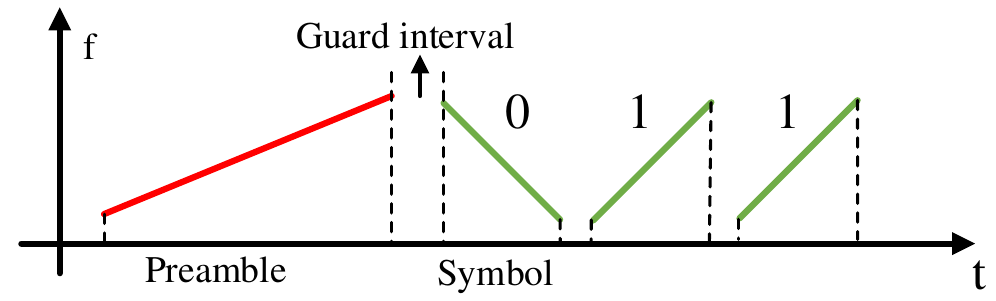}\\
  \caption{An example CSS frame}
  \label{fig:css frame}
\end{figure}

\paragraph{CSS} 
With the ability to handle dynamic channel conditions and extend communication ranges, CSS is arguably the most dominant modulation technique for acoustic communications~\cite{ChirpCommunication,ALPS,AALTS,ALPSPre}.
% becomes the most commonly used modulation technique. 
% CSS is a known technique for LoRaWAN~\cite{LoRa} (IEEE 802.15.4a) that aims to achieve long-range communication with low power consumption. 
It employs noise-resilient chirp signals as carriers,  making it robust to co-channel interference, path losses, multipath fading, and the Doppler effects~\cite{ChirpCommunication}. 
%These salient features make CSS and its variants adaptive to dynamic channel conditions and achieve a long communication range, at the cost of transmission rate. 
% In addition, it is suitable for mobile scenarios as chirp carriers are also free from the Doppler effect. %These salient properties make CSS target for relatively long-range communication than FSK and OFDM where
A CSS frame starts with a preamble followed by different symbols as illustrated in Fig.~\ref{fig:css frame}. Both preambles and data symbols in a CSS frame use chirp signals but with different chirp rates. 
%The way of mapping chirp rates to bit streams is the fundamental issue in CSS modulation. 
%An example of a Binary Orthogonal Keying (BOK) CSS frame is illustrated in Fig.~\ref{fig:css frame}. %The preamble is used for synchronization and the data symbols are used to deliver messages. Often, guard intervals are inserted between a preamble and data symbols to mitigate Inter-Symbol-Interference (ISI).
Though with significantly improved robustness, CSS still suffers from low transmission rates.

\subsubsection{Challenges for Waveform Design} 
Next we discuss two design challenges in devising or selecting suitable waveforms for acoustic sensing and communication. 
\begin{itemize}
    \item % As we mentioned earlier in this section, the available 24~\!kHz acoustic bandwidth has \emph{audible} and \emph{inaudible} frequency parts. 
    Though transmissions in the range from 20Hz to 18KHz enjoy better channel gains and a relatively large bandwidth,
    % hence can contribute to better system performance. However, this audible part 
    their audibility limits the application scenarios. 
    % hence are less attractive. 
    In contrast, transmissions in the inaudible range though in absence of unintended noise, suffer from a limited bandwidth and significant  attenuation. Therefore, in choosing a suitable bandwidth for target applications, one needs to consider,
 \begin{challenge}
 \label{clg:bw_tradeoff}
 Trade-offs among channel quality, bandwidth and audibility.
 \end{challenge} 
 %   This challenge, together with \textit{\textbf{Challenge-III}}, are both relevant to transmission-induced noise issue, so they require application-specific signal processing techniques to handle, as further discussed in Section~\ref{sec:biometric sensing} and~\ref{sec:future direction for channel characteristics}.

    %

    %
    \item % As we outlined in previous content, to achieve satisfactory throughput in aerial acoustic communication, we should narrow down the communication range and adopt efficient modulation techniques such as FSK or OFDM. And to achieve reliable long-range communication, 
    As discussed in TABLE~\ref{tab:waveform design comparision for communication}, existing waveforms either suffer from short communication ranges or low data rates. 
    
\begin{challenge}
\label{clg:comm_range}
 Designing modulation schemes that are suitable for both {high-speed and long-range inaudible aerial communication.}
\end{challenge}

To do so, one needs to devise robust high-order modulation techniques.  One such approach is presented in Section~\ref{sec:long range high speed communication via loose orthogonal modulation}.
\end{itemize}

\subsection{Temporal Feature Extraction}
\label{sec:temporal features}
%
% Temporal features are often related to the time information of a particular acoustic waveform.
Temporal features refer to timing information such as onsets, time-of-arrival and time-difference-of-arrival. This section focuses on the techniques to extract these features. 
%First, we introduce two correlation-based techniques, namely, one-way and two-way sensing, and then present phase-based methods that achieve fine-grained time information. }
%As all these time information extraction techniques heavily depend on the correct onset point detection of a particular reference signal, we thus start with onset point detection.

\begin{figure}[t]
    \centering
    \includegraphics[width=0.95\columnwidth]{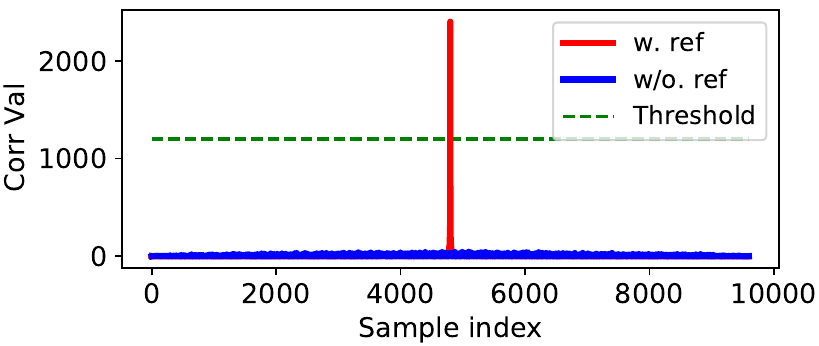}
    \caption{Illustration of signal onset detection.}
    \label{fig:signal start detection}
\end{figure}

\begin{figure}[b]
    \centering
     \subfigure[Near-far effect.]
    {
        \label{fig:near far effect}
        \includegraphics[width=0.45\columnwidth]{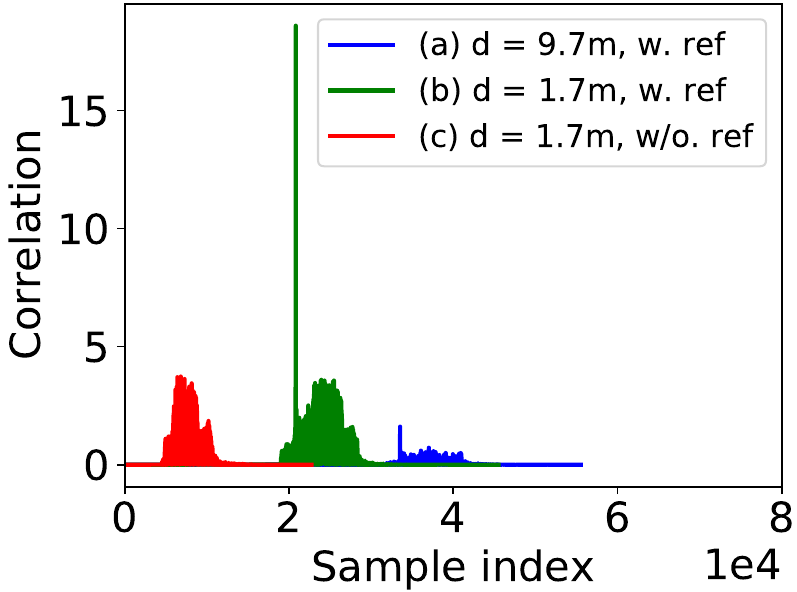}
    }
    \hspace{0.005\linewidth}
    \subfigure[Clap noise and multipath effect.]
    {
        \label{fig:multipath effect}
        \includegraphics[width=0.45\columnwidth]{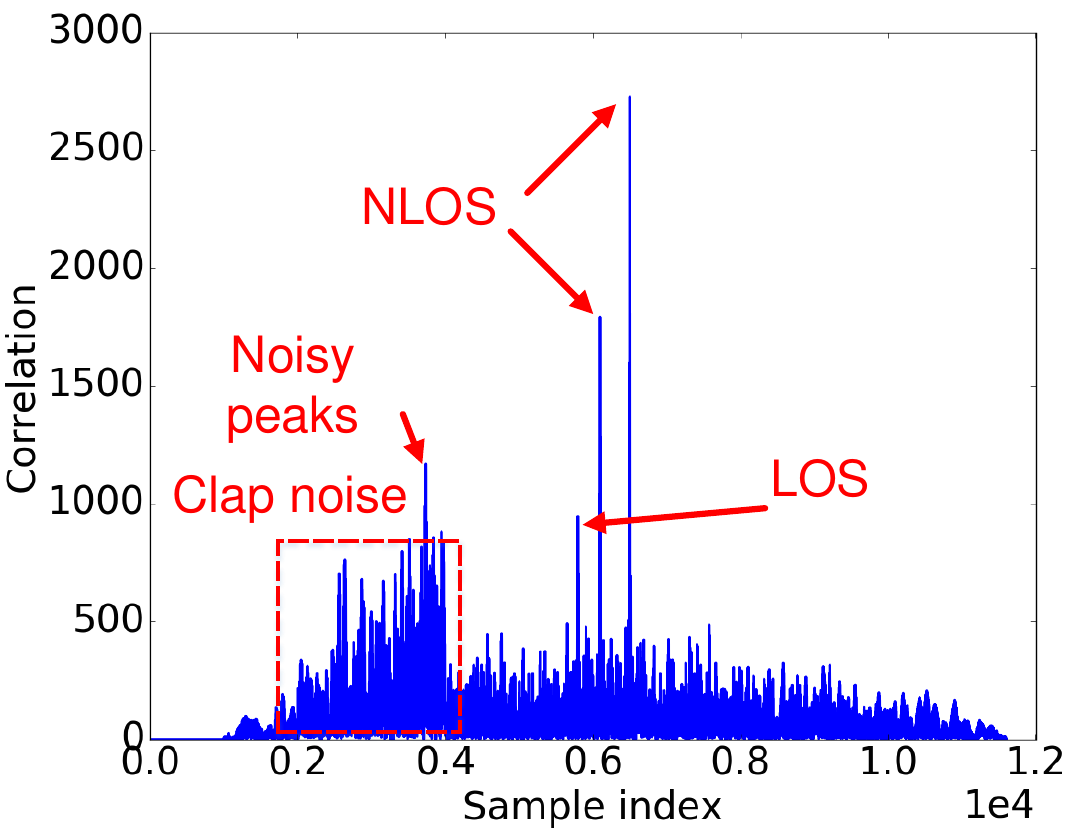}
    }
\caption{(a) The correlation peak of distant samples is even weaker than the noisy peaks from closer ones, making it challenging to set an appropriate threshold for reliable onset detection, (b) Interference including clap noise and multipath effect can lead to incorrect timestamp estimation.}
\label{fig:preamble detection}
\end{figure}

\subsubsection{Onset  Detection}
\label{sec:onset point detection}

Onset detection, determining the presence and the start time of a particular reference signal\footnote{In communication system, the reference signal is often the preamble and hence onset detection is the same as preamble detection.} is the cornerstone of acoustic communication and time-sensitive sensing applications. %In this section, we outline the basics of onset point detection.
 
%\paragraph{Basics}
%Onset point detection contains the tasks of both determining the presence of a particular reference signal and its arrival time. 
Onset detection is often achieved via cross-correlation between the captured signal and a known \textit{reference signal}. The waveform of the reference signal is carefully designed so that there exists a sharp peak in the correlation results when the reference signal is present; otherwise, the maximum correlation value is far below a pre-defined threshold (Fig.~\ref{fig:signal start detection}). As a result, in absence of multipath effects or strong interference, a simple threshold-based approach is sufficient to detect the presence of the reference signal and the timestamp of the maximum peak is recorded as the onset of the signal.

In practice, the accuracy of threshold-based onset detection is degraded by three phenomena, namely, device heterogeneity, the  ``near-far'' effect and the multipath effects. Device heterogeneity refers to the fact speakers and microphones on different devices have different gains. A constant threshold may not work for all devices. 
The ``near-far'' effect, a term originated from wireless communication systems, describes the phenomenon where the signal power received at a base station is dominated by the signals from closer user devices due to signal attenuation over distance. 
%This channel dynamic is notably detrimental. Specifically, 
Similarly, in acoustic systems, when a transmitter and a receiver are close, the resulting correlation peak at the receiver can readily exceed a pre-defined threshold and hence the reference signal is detected (Figure~\ref{fig:near far effect}(a)). However, when their  distance becomes larger, the respective correlation peak may fall below the threshold and in fact sometimes below the values caused by close-by interfering sources. In a multipath rich environment, a receiver not only captures the Line-of-Sight (LoS) signal but also receives multiple delayed and attenuated copies. These delayed and attenuated copies, called None-Line-of-Sight (NLoS) signals, can add up constructively and thus surpass LoS signals in peak intensity. A naive thresholding approach or simply finding the largest peak will result in false onset detection.

Fig.~\ref{fig:near far effect}(b) illustrates the effects of the three phenomena when the reference signal is a chirp signal. In the experiment, hand claps prior to the transmission of the chirp signal result in noticeable peaks even after cross-correlation at the receiver. Peaks due to NLOS paths are in fact larger than that from the LOS path.

\subsubsection{Timing Estimation}
\label{sec:basic timing measurements}
Estimating Time-of-Arrival (ToA) or Time-Difference-of-Arrival (TDoA) of transmitted acoustic waves at target receivers builds upon onset detection and plays an important role in ranging and localization. Depending on whether the intended targets are equipped with acoustic transceivers, existing techniques can be further categorized into device-based and device-free approaches. Among device-based approaches, \textit{one-way} or \textit{two-way} sensing methods typically utilize cross-correlation to achieve sample-level timing resolutions. In device-free approaches, to estimate the time-of-flight between transmission and the reception of reflected waves from a target, phase information can also be used to achieve  subsample-level resolutions. %XXX: I still disagree with the assertion of subsample-leve resolution
%[CC]: To be done

%The time information dependent on starting pointing detection
%can be generally classified into correlation based approaches including \textit{one-way} or \textit{two-way} sensing paradigms that exhibit sample-level resolution, and \textit{phase-based} methods that achieve subsample-level resolution. And the time information here often refer to Time-of-Arrival (ToA) or Time-Difference-of-Arrival (TDoA). 

\begin{figure}[b]
    \centering
    \subfigure[ToA via one-way sensing.]
    {
        \label{fig:toa via one way}
        \includegraphics[width=0.45\columnwidth]{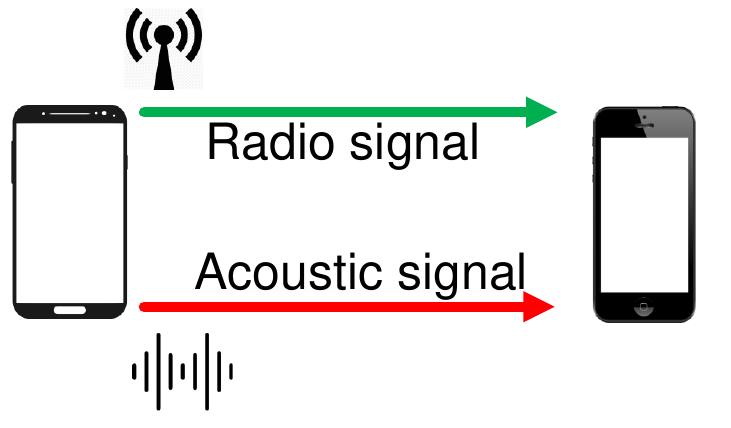}
    }
    \hspace{0.002\textwidth}
    \subfigure[TDoA via one-way sensing.]
    {
        \label{fig:tdoa via one way}
        \includegraphics[width=0.45\columnwidth]{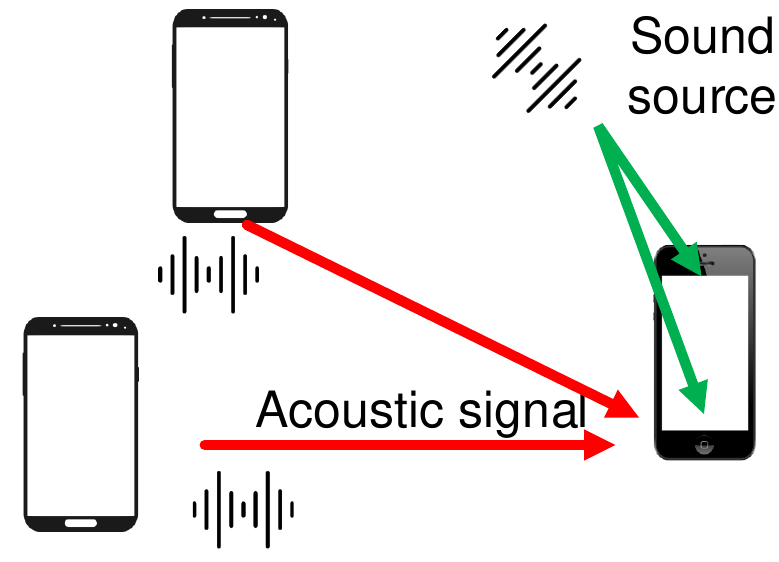}
    }
\caption{One-way sensing paradigm.}
\label{fig:one way sensing paradigm}
\end{figure}

\begin{figure}[t]
    \centering
    \subfigure[ToA via two-way sensing.]
    {
        \label{fig:toa via two way}
        \includegraphics[width=0.45\columnwidth]{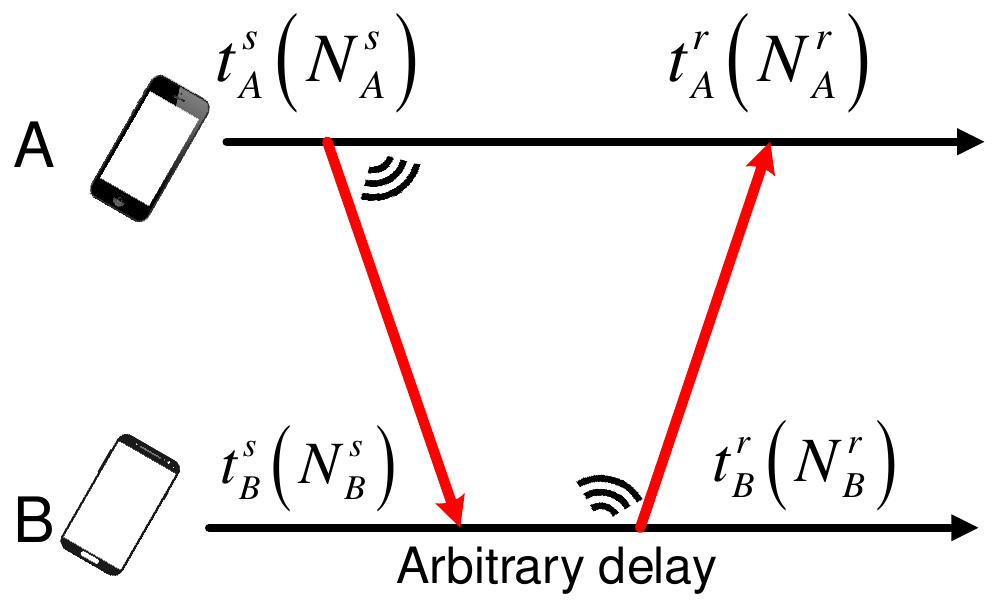}
    }
    \hspace{0.002\textwidth}
    \subfigure[TDoA via two-way sensing.]
    {
        \label{fig:tdoa via two way}
        \includegraphics[width=0.45\columnwidth]{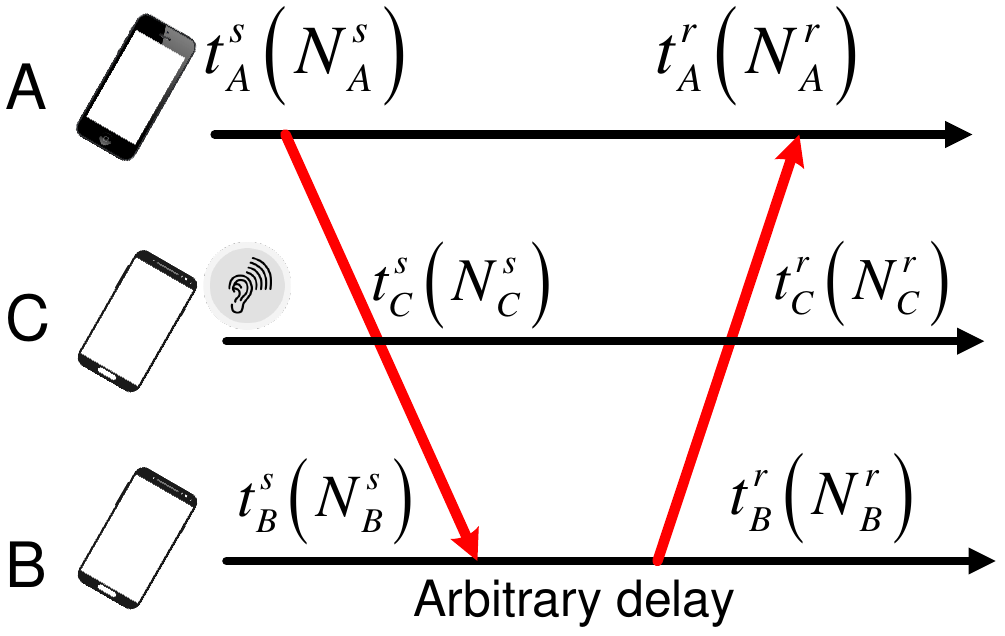}
    }
\caption{Two-way sensing paradigm.}
\label{fig:two way sensing paradigm}
\end{figure}

\paragraph{One-Way Sensing} One-way sensing generally refers to a sensing paradigm where time information is obtained through unidirectional transmissions. In this paradigm, the transmitter or receiver can be multiple separated devices or a single device with multiple channels as shown in Fig.~\ref{fig:one way sensing paradigm}. One-way sensing is achieved via tight synchronization so that a receiver knows precisely the onset of an acoustic transmission to estimate its flight time.
%It estimates ToA or TDoA through one-shot directional transmissions. 
For time synchronization, this approach often exploits another high-speed signal sources~\cite{RFBeep,Guoguo}. These signal sources are often radio signals such as WiFi, Bluetooth, and Zigbee whose propagation time is negligible within the maximum range of acoustic waves. For ToA estimation, an acoustic signal and a synchronization signal are transmitted simultaneously. A receiver determines the ToA from the difference in arrival time between the two signal sources. %In this approach, timing information can be obtained without any coordination. 

For TDoA estimation, multiple transmitters or receivers need to be tightly synchronized. In some cases, transmitters or receivers co-locate physically on a single device as shown in Fig.~\ref{fig:tdoa via one way}. In transmitter-synchronized systems~\cite{ALPSPre,ALPS}, acoustic transmissions are triggered concurrently and they arrive at a receiver after different propagation delays. TDoA is then obtained by cross-correlating received samples with known reference signals. %The timing information, namely the delay among each arrival signals, is determined by the gap between multiple peaks in the correlation results.
In receiver-synchronized systems, TDoA is computed by cross-correlating the received samples from different receive channels~\cite{SnoopingKeystrokes} as shown in Fig.~\ref{fig:tdoa via one way}. 
%Note that in this case, both passive and active sound can be used. 
%XXX: reference?
%[CC]: Done
%A special case for one-way sensing is radar wherein both the transmitter and and receiver are physically on a same device hence are in-phase. In this case, the temporal information is the delay between self-recorded signal and desired reflections. 
One-way sensing is simple and effective in extracting time information.
However, the main drawback lies in its needs for tight synchronization, especially for cases where distributed devices are involved. This tight synchronization requirement can be compromised by {\bf Challenge~\ref{clg:system_delay}} (i.e., uncertain system delays), thereby significantly affecting the timing performance. 
Though the uncertainty can be mitigated by either kernel-level implementation on general-purpose OSs or customized hardware, the applicability of one-way sensing is severely restricted. 

\paragraph{Two-Way Sensing} %Two-way sensing extracts time information in a synchronization-free manner at the cost of both extra hardware and processing complexity. 
To relax the need for tight synchronization, two-way sensing has been proposed at the expense of increased hardware and processing complexity.
In two-way sensing, acoustic transmissions are bi-directional. It is assumed that each device is equipped with both a speaker and a microphone. Extracting ToA information between two devices is shown in  Fig.~\ref{fig:toa via two way}. At time ${t_A^s}$, device A starts an acoustic transmission. Device B detects the acoustic signal at time ${t_B^r}$ and starts another transmission at time ${t_B^s}$ after an arbitrary delay. Device A detects the second transmission at time ${t_A^r}$. Therefe time information (ToA) can be derived as~\cite{BeepBeep}:
\begin{equation}\label{eq:toa via two way}
{t } = \frac{1}{2}\left( {t_A^r - t_A^s} \right) - \frac{1}{2}\left( {t_B^r - t_B^s} \right).
\end{equation}
If both transmissions can be received by a third device, then time differential information (TDoA depicted in Fig.~\ref{fig:tdoa via two way}) can be derived by~\cite{ARABIS}:
\begin{equation}\label{eq:tdoa via two way}
{t } = \frac{1}{2}\left( {t_A^r - t_A^s} \right) + \frac{1}{2}\left( {t_B^r - t_B^s} \right) - \left( {t_C^r - t_C^s} \right).
\end{equation}

Note that in presence of uncertain system delay ({\bf Challenge~\ref{clg:system_delay}}),  ${t_A^s,}$ ${t_A^r, t_B^r, t_B^s, t_C^r, t_C^s}$ cannot be recorded precisely in user applications. To overcome such a challenge, two novel techniques have been proposed in \cite{BeepBeep}. First, in addition to transmissions from other devices, each transmitting device also records its own transmission through its on-board microphone. Second, sample counting in the audio buffer of a device is used to estimate the time elapsed between consecutive acoustic receptions. Combining with the known distance (and thus known propagation delay) between the microphone and the speaker on each device, one can then estimate TOA and TODA following Eqn.~\eqref{eq:toa via two way} and ~\eqref{eq:tdoa via two way}.

\subsubsection{Phase-enabled Fine-grained Timing} % via Phase Information}
\label{sec:phase-enabled accurate timing}
Both one-way and two-way sensing for TOA or TDOA rely on cross-correlation to determine the onset of signals, whose resolution is limited by the sampling rate of respective devices. At a sampling rate of ${f_s = 48}$~\!kHz,  the time resolution is upper bounded by $\frac{1}{f_s} \approx 2.1 \times 10^{-5}$~\!s, or equivalently a range resolution of 7~\!mm if the sound speed is 340~\!m/s. However, CC-based onset detection often suffers from 2 to 3 samples errors~\cite{FingerIO}, resulting a timing error from 40 to 60 $\mu$s.To achieve finer time granularity, other signal features should be exploited. 
In device-free sensing, since the acoustic transmitter and receiver are co-located, to estimate the time flight of reflected waves from interested targets, phase information can be exploited. The relationship between phase changes and time is waveform-dependent. 
%
%To break the resolution limit constrained by the sampling rate in CC-based methods so as to achieve finer time granularity, phase-based approaches have been proposed.  
%Phase-based sensing techniques can achieve finer time information, breaking the resolution limit constrained by the sampling rate. 
%Despite the waveform, phase is a continuous variable other than the discrete sample count in CC-based methods hence is more sensitive and reflect finer time resolution. %Often, the absolute phase value is not useful but the phase difference is more meaningful. Meanwhile, extracting phase difference also adds immunity to the uncertainty system delay, making phase-enabled timing superior to basic time mechanism.
We next present the basic approaches to extract phase from three waveforms discussed in Section~\ref{sec:waveform for sensing}, namely, pure tone signals, FHSS signals, and chirp signals. 

\begin{figure}[b]
  \centering
  \includegraphics[width=2.5in]{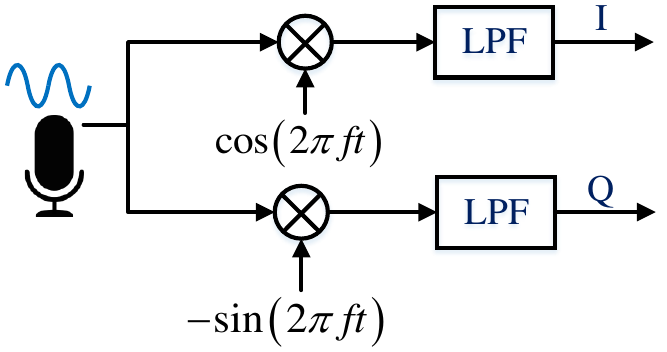}\\
  \caption{Structure of coherent receiver. LPF in figure denotes Low Pass Filter.}
  \label{fig:coherent receiver}
\end{figure} 

\begin{figure}[thp]
    \centering
     \subfigure[Time to phase via pure tone and FHSS signals.]
    {
        \label{fig:phase for pure tone and FHSS}
        \includegraphics[width=0.85\columnwidth]{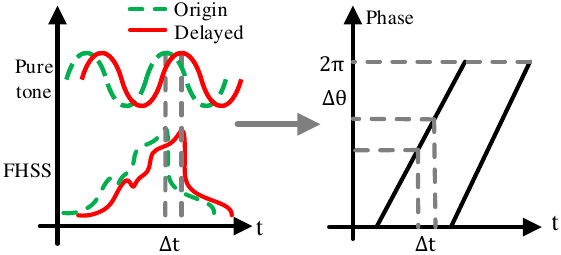}
    }
    \\
    \subfigure[Time to frequency via chirp mixing.]
    {
        \label{fig:phase for chirp}
        \includegraphics[width=0.85\columnwidth]{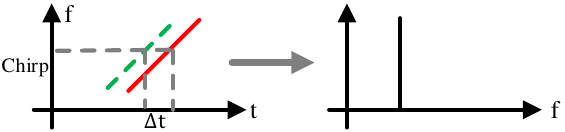}
    }
\caption{(a) Temporal characteristics are converted to phase in pure tone and FHSS signal based system. (b) Chirp mixing transforms time measurements into frequency information.}
\label{fig:phase-based method}
\end{figure}

\paragraph{Pure Tone Signals}
The phase of pure tone signals can be extracted by a coherent receiver whose structure is shown in Fig.~\ref{fig:coherent receiver}. In a coherent receiver, two identical copies of an input signal are multiplied by ${\cos \left( {2\pi ft} \right)}$ and its ${\frac{\pi}{2}}$ phase shifted version ${- \sin \left( {2\pi ft} \right)}$, respectively. After a low pass filter, the In-phase (I) and Quadrature-phase (Q) components can be obtained. The absolute phase $\theta$ in $s\left(t\right) = \cos\left( 2\pi ft + \phi \right)$ is calculated by $\tan^{-1}(Q/I)$. Phase changes can be determined by subtracting consecutive absolute phase as illustrated in Fig.~\ref{fig:phase for pure tone and FHSS}. 
Consider a pure tone oscillating at ${f_c = 20}$ kHz and the sampling rate is ${f_s = 48}$ kHz. One sample interval corresponds to a phase change of $\frac{f_c}{f_s}\times 2\pi = \frac{5}{6}\pi$. Therefore, if the detectable phase change is below $\frac{5}{6}\pi$, one can readily achieve sub-sample time resolution. Since such a $\frac{5}{6}\pi$ phase change is comparable to the maximum $2\pi$ value, phase-based approaches generally outperform CC-based ones in time granularity. A major shortcoming of pure tone signals is their vulnerability to background noise and multipath effects.

%XXX: separate pure tone and FHSS signals
%[CC]: Do not quite catch up how to separate. I realize that pure tone and FHSS signals are without labels. So I add labels for them. If separate them, the right part is not easy to handle. 

%\textcolor{green}{To extract phase change via pure tone signals is often achieve by coherent receiver and more about this can be found in Section~\ref{}. FHSS signals and chirps can also reflect phase information and the way to extract phase information from these two signals can be found in Section~\ref{}}.
%Signal mixing is another approach to achieve finer time resolution and this method is only suitable for chirp waveforms. This approach transforms the time delay information between two chirps to frequency estimation. As frequency estimation can also include phase part, this manner can readily break the resolution limit of CC-based approaches. \textcolor{green}{More discussions on this can be found in Section~\ref{}.}

\paragraph{FHSS Signals} 
%As we have already mentioned in Section~\ref{sec:waveform for sensing}, FHSS signals have better auto-correlation properties than pure tone signals hence gain the ability to distinguish multipath signals. 
%The basic steps to retrieve phase changes in FHSS signals are listed as follows. First, an orthogonal sequence such as GSM training sequence, barker code, M-sequence, or ZC sequence is chosen as the baseband signal. Then, this baseband signal is up-converted by a pure tone carrier. Following that, up-sampling is performed. At this time, the output signal from above processes can be utilized for sensing. However, if the baseband signal is in frequency domain before up-conversion, an extra Inverse Discrete Fourier Transform (IDFT) operation is required. At the receiver side, the aforementioned processes are reverted so as to obtain the basedband signal. 
The demodulated baseband signal $y(n)$ from receiver together with the original known one $x(n)$ are then utilized to estimate the Channel State Information (CSI, or Channel Impulse Response, hereafter, we will use CSI instead) $h(n) = \frac{y(n)}{x(n)}$ (in complex form), where $n$ represents the sample index. We are often particularly interested in a specific index $n_s$ of $h(n)$ as they may reflect interactive actions such as finger movements or respiration. The absolute phase of $h(n_s)$ is not useful but if we track its phase difference in consecutive frames as illustrated in Fig.~\ref{fig:phase for pure tone and FHSS}, we can be able to track the phase of, say finger-generated echo hence locate it. 
%Finally, the phase change can be extracted from $h_d(n) = h(n) - h(n-1)$ as illustrated in Fig.~\ref{fig:phase for pure tone and FHSS}. 
This can also handle ambient static noises as they are excluded by subtraction. The aforementioned processing steps including CSI estimation first and then CSI phase differentiation has become the routine for FHSS signal based applications. 
FHSS signal is sensitive enough to detect small phase changes but subjects to various channel distortions such as path loss and the Doppler effect hence is only suitable for around device sensing. To sense at long range or low SNR, chirp signals are more appropriate. 
%XXX: there is no need to discuss how FHSS is generated. You only need to focus on the receiver side. It is still not clear how phase is extracted from  channel impulse response. please rewrite.
%[CC]: Done
\paragraph{Chirp Signals} Extracting phases from chirp signals is achieved through chirp mixing, which converts time information into frequency features as illustrated in Fig.~\ref{fig:phase for chirp}.
Assume that the received signal $s'(t)$ is a delayed and attenuated version of a  transmitted signal $s(t) = \cos\left(2\pi \left(  f_\mathrm{min} t + \frac{B}{2T}t^2 \right) \right)$. In other word, $s'(t) = As(t - \Delta t) = A\cos\left(2\pi \left(  f_\mathrm{min} \left(t- \Delta t \right) + \frac{B}{2T}\left(t - \Delta t\right)^2 \right) \right)$. In chirp mixing, we  multiply $s(t)$ and $s'(t)$ and feed the resulting signal through a low pass filter. The mixed results become $s_\mathrm{mix}(t) = A\cos\left(2\pi \Delta t \frac{B}{T} t + 2\pi f_\mathrm{min} \Delta t - \pi\frac{B}{T}\Delta t ^ 2 \right)$. Clearly, $s_\mathrm{mix}(t)$ is a single tone signal with frequency $ \frac{B\Delta t}{T}$. Applying a Discrete Fourier Transform (DFT) on  $s_\mathrm{mix}(t)$, we can estimate $\Delta t$ by $f_p\cdot T/B$, where $f_p$ is the peak frequency after DFT. From the discussion, we see that the timing resolution of chirp mixing is upper bounded by its frequency resolution multiplied by a constant. The frequency resolution of DFT is proportional to $\frac{1}{T}$. Therefore, the granularity of $\Delta t$ is proportional to $\frac{1}{B}$. For example, for a chirp signal in the range of 18KHz to 24KHz, we have a timing resolution of 0.167ms. To further improve the timing resolution, phase information in the mixed signal can be utilized. We see that in the frequency bin $f_p = \Delta t \frac{B}{T}$, the phase of the mixed signal $\phi(t) = 2\pi \left( f_\mathrm{min} \Delta t - \frac{1}{2}\frac{B}{T}\Delta t ^ 2\right)$. Therefore, a time difference of 10$\mu$s corresponds to a phase difference of 0.18 radian when $f_\mathrm{min} = 18$~\!kHz, notable enough to be detected. The phase differences can be estimated over multiple chirps. 
%XXX: double check. not clear what is the limit of phase resolution. 
%[CC]: Done

% as $B$ and $T$ are two constants. The DFT results contains both frequency and phase component, making $\Delta t$ a continuous variable \textcolor{red}{hence better resolution than CC-based methods that are in discrete format and often suffer from 2 to 3 samples errors~\cite{FingerIO}. }
%Given that CC-based methods, regardless of the utilized waveform, often suffer from 2 to 3 samples errors~\cite{FingerIO}, phase-based approaches could readily achieve finer time resolution. 
Theoretically, phase-based methods can be combined with one-way sensing since the transmitter and receiver devices are synchronized. However, carrier frequency offsets~\cite{CAT} due to clock drifts, introduce constant phase shifts and result in large timing errors in the long run.

\subsubsection{Challenges for Temporal Feature Extraction} 
\begin{itemize}
    \item As discussed in Section~\ref{sec:onset point detection}, the near-far effect and device heterogeneity problem make the use of constant threshold in onset detection inadequate. Multipath effects and strong interference also pose challenges in onset detection. 
    \begin{challenge}\label{clg:robust onset detection}
    Robust onset detection under dynamic channel conditions.
    \end{challenge} 
    A possible solution is to set thresholds dynamically based on the perceived noise level. More discussions can be found in Section~\ref{sec:localization}.
    %
    % \item Current strategies in extracting timing information generally have to trade-off complexity with  have to strike a balance between two desirable properties, namely low-complexity processing (e.g., in one-way sensing manner) and synchronization-free nature (e.g., in two-way sensing approach). It is indeed challenging in:
    % \begin{challenge}
    % Designing an appropriate approach that has low complexity and is synchronization free.
    % \end{challenge}
  
    % %
    % Exploiting special time-associated properties could be a solution for certain applications; relevant discussions can be found in Section~\ref{sec:future direction for temporal features}.
    %
    % \item When involving multiple heterogeneous devices, the sampling frequency offset caused by clock difference can lead to large integration errors in the long run, especially for one-way sensing approach aiming to extract fine-grained temporal features. Therefore, it is a challenge in:
    % %
    % \begin{challenge}
    % Obtaining fine-grained temporal characteristics in the long-run.
    % \end{challenge}
    % Discovering the law of the impacts from this sample frequency offset on the final temporal characteristics and designing appropriate compensation techniques would be a potential method; details on this can be found in Section~\ref{sec:tracking}.
    \item Design considerations such as waveforms, duration and bandwidth of signals  will affect the accuracy and resolution of timing estimation in different approaches. In general, longer time interval and larger bandwidth lead to higher timing accuracy. Transmitting multiple frames is also instrumental  phase estimation. However, there is a trade-off between the timeliness of the estimation (e.g., in highly dynamic scenarios) and performance. 
\begin{challenge}\label{clg:timing estimation trade-off}
Identifying suitable configurations to achieve a good trade-off between performance and adaptiveness to dynamic changes in timing estimation.
\end{challenge}
\end{itemize}

\subsection{Channel Profiling}
\label{sec:channel characteristics}
\begin{figure}[t]
  \centering
  \includegraphics[width=3.5in]{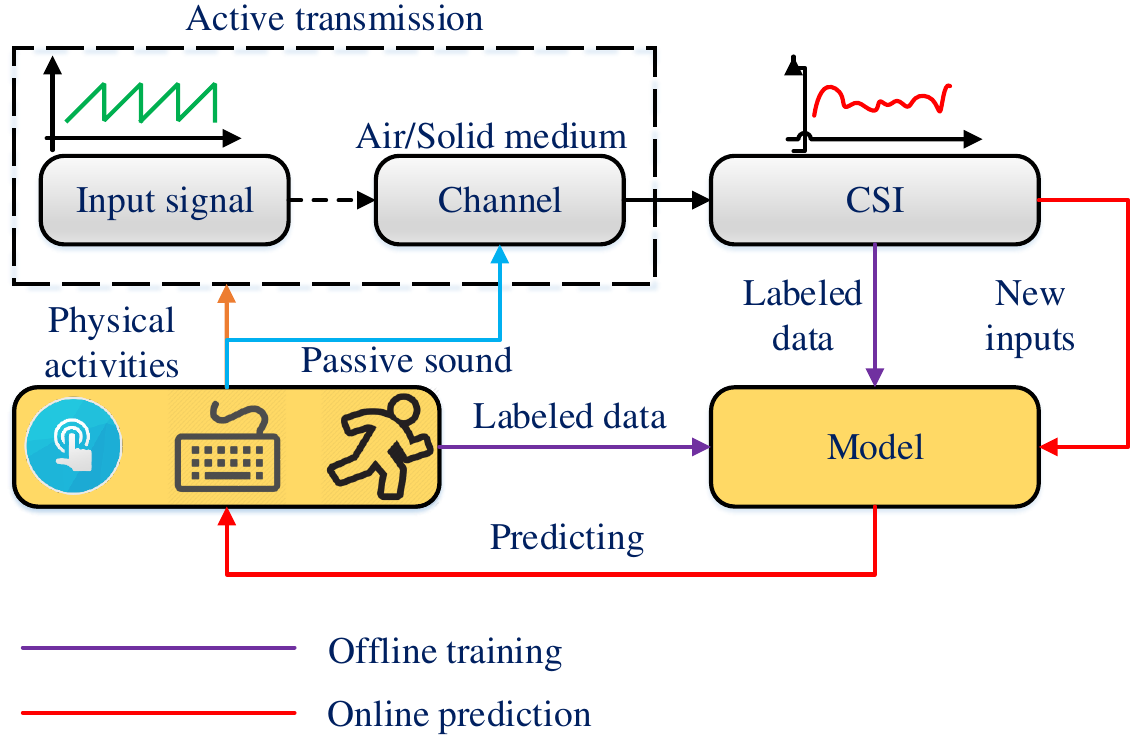}\\
  \caption{Canonical data flow of common channel modeling methods.} %\textcolor{green}{[Jun] Why this figure appears here but not cited here?}}
  %XXX remove passive sound
  %[CC]: I believe passive sound is also feasible. 
  \label{fig:channel modeling}
\end{figure}

%
%\subsubsection{Non-Temporal Feature Extraction}
% 
This section presents channel characterization techniques for application development. The objective of channel profiling is to find specific acoustic features and their precise relationships with respective states. These acoustic features, formally defined as channel characteristics or Channel State Information (CSI), here includes influences from both acoustic front-ends and acoustic medium that shape these acoustic features. The construction of a precise relationship is also called channel modeling. In can be in the format of explicit mathematical model or implicit mapping, derived from model-driven and data-driven approaches, respectively. A canonical data flow of common channel modeling methods are shown in Fig.~\ref{fig:channel modeling}. We now first start with CSI representation. %In this section, we first discuss CSI representations, followed by channel modeling. }
%We first discuss channel properties, then elaborate on Channel Impulse Response (CSI) representation, and finally talk about channel model construction. 

%\subsubsection{Channel Property} 
%\label{sec:channel property}

%[CC]: we need to wait for PACE being accepted 

%These two signals have sharp difference in propagation speed where the speed of AS is 340~\!m/s (at temperature of 25$^\circ$) and for SS, the speed would be over 2000~\!m/s (depending on the medium property). 

\subsubsection{CSI Representation}
\label{sec:cir representation}
As we mentioned before, CSI representation is to find a specific acoustic feature, say Doppler frequency, to decipher respective states, say hand gestures. 
To find a successful CSI representation is non-trivial and often requires sophisticated domain knowledge and intensive observations. 
The first thing in CSI representation would be acquiring adequate preliminary information on common acoustic features and acoustic channel characteristics, which can be found in Section~\ref{sec:acoustic hardware}. Having profound knowledge on acoustic basics would be helpful to strike best features to reflect respective states. 
In this section, we present typical CSI representations for acoustic characteristics outlined in Section~\ref{sec:acoustic channel property}.

\textbf{Path Loss, Reflection, Refraction, and Diffraction:} These generic wireless properties of acoustic signals can be a double-edge sword for acoustic sensing. They can on one hand attenuate signal strength and result in multipath effect hence affect signal quality and therefore are harmful; on the other hand, they can reflect spatial information and therefore is useful for applications such as localization. A common adoption to reflect these properties is using correlation spectra. To avoid the side effects by these properties, one can use sophisticated waveform design outlined in Section~\ref{sec:waveform design} to obtain correlation spectra that can disentangle multiple reflections. Often, waveforms such as chirp signals that have high correlation  and processing gain is favorable as multiple reflections are separable if using these signals. For passive sensing system, the features in the correlation spectra may not be noticeable hence require dedicated signal processing. Since these signal processing tricks are application-dependent, we postpone any further discussions on the application layer. To harness the multipath effect, one should adopt waveforms that can enrich acoustic profiles. This could be achieved by wide-band sensing using signals of a relatively long duration. However, using signals of long duration would increase latency hence trade-off between latency and duration should be made. 

\textbf{Doppler Effect:} To estimate the Doppler effect of a target, one often lets a device to transmit a pure tone signal due to its high Doppler accuracy as we mentioned in Section~\ref{sec:waveform for sensing}. If the target owns an acoustic-enabled device, we then use this device to capture transmitted signals and perform DFT analysis in a sliding window fashion. Through this, we obtain the major frequency of received signals over time. To this end, we follow Eqn.~\ref{eq:doppler effect} to estimate the Doppler effect. Note that in practice, multiple tones may be transmitted and one can average the result from each tone so as to improve estimation accuracy. For targets without any devices, we can track the Doppler frequency by their acoustic reflections. The major difficulty here lies in the method to pin down target induced acoustic reflection in the presence of strong self-interference. The adoption of self-cancellation techniques and the utilization of prior information would be common practice. For instance, the walking speed of human beings typically being around 1~\!m/s can be used to confine the search range in DFT results to obtain a target's moving speed. 

\textbf{Temperature Effect:} As we mentioned earlier in Section~\ref{sec:acoustic channel property}, the propagation speed of acoustic signals is dependent on ambient temperature. Therefore, CSI representation in this case is rather straightforward. Since speed can be derived by the division between distance and time, and we have plentiful approaches for timing estimation discussed in Section~\ref{sec:basic timing measurements}, time is thus more suitable for CSI representation. 

\textbf{Acoustic Dispersion:} Recalled that acoustic dispersion describe a phenomenon that different acoustic frequency components propagate at heterogeneous speeds in solid medium as revealed by Eqn.~\ref{eq:acoustic dispersion equation}. This phenomenon could be observed by a rather distinctive time-domain waveform dictated in Fig.~\ref{fig:acoustic dispersion}. This in turn indicates that time-domain acoustic waveform would be a possible choice for CSI representation. Application based on time domain acoustic dispersive waveform can be found in Section~\ref{sec:solution or application}. Meanwhile, as different frequency components arrival at different time, when converting time-domain waveform into frequency spectrogram, we may observe particular waveform shaped by frequency versus time. If we could interpolate this waveform by, say linear interpolation, we can thus utilize the curvature of the waveform for CSI representation. This curvature may reveal distance~\cite{UbiTap} or material information. When the curvature is not obvious or the waveform cannot be readily interpolated, magnitude spectrogram or other common acoustic features such as MFCC or GFCC may be alternative choices. Note that acoustic dispersion phenomenon is observable only when there are multiple frequency components so that in active sensing, signals of wide bandwidth like chirp signal is preferable. In passive sensing, acoustic signals produced by natural events generally contain a rich set of multiple frequency components.

\textbf{Acoustic Resonance:} 
In acoustic resonance enabled applications, %we should also utilize acoustic signals with wide bandwidth so that we can observe the highest channel gain at a particular frequency. In this particular channel characterization method, 
the goal is to inspect resonance properties of a particular object. Therefore, the primary setting is to place sensors including (piezo) microphone and (piezo) speakers tightly on this object and focus on the properties of structure-borne acoustics. We then actively transmit wide-band signals such as chirp signals through this piezo speaker and let the piezo microphone to capture the transmitted signals. 
Following that, DFT is typically applied to check the frequency response of received signals. In order to detect minor resonance frequency change, it is desirable to utilize a high sampling rate so that DFT results can achieve better frequency resolution. From the aforementioned processing steps, we can infer possible CSI representations, for instance, peak frequency or frequency response. An appropriate CSI representation can often ease significant efforts on model construction. 
With appropriate CSI representations, the next step is to build an application-specific model.

\subsubsection{Channel Model Construction}
\label{sec:channel model construction}

Constructing an appropriate channel model is the cornerstone for acoustic sensing application development. A canonical data flow of common channel modeling methods are shown in Fig.~\ref{fig:channel modeling}. 
The basic idea behind channel model construction is to map respective states, say hand moving directions, to certain CSI representations, say Doppler frequency. A naive approach, also called \textit{Data-Driven} (DD), to construct this model is through direct mapping, which often involves massive data and intensive training. Another method, known as \textit{Model-Driven} (MD), formulates this mapping with a close-formed expression, which are often more efficient but non-trivial to achieve. The respective advantages, disadvantages, and suitable applications for these two methods are displayed in TABLE~\ref{tab:processing technique comparison}. In this section, we highlight the basic steps for aforementioned two common channel model construction methods.

%If the mapping can be formulated with a close-formed expression, we then call it \textit{Parametric Modeling} (PM); otherwise we call it \textit{Data-Driven} (DD) approach. The respective advantages, disadvantages, and suitable applications for these two methods are displayed in TABLE~\ref{tab:processing technique comparison}. 

\begin{table}[b]
\caption{Comparison of MD and DD approaches.}
\label{tab:processing technique comparison}
\centering
\begin{tabular}{|c|c|c|c|c|}
\hline
 & \textbf{Problem} & \textbf{Advantages} & \textbf{Disadvantages} & \textbf{\makecell{Suitable\\applications}}\\ \hline
MD & \makecell{Regression\\ problem} & \makecell{Effective\\and\\efficiency} & \makecell{Require domain\\knowledge,\\time-consuming} &\makecell{Temporal\\feature\\/channel\\characteristics}\\ \hline
DD & \makecell{Categorical\\ problem} & \makecell{Simple\\in\\model\\designs} & \makecell{Massive data,\\computation-\\intensive,\\heterogeniety\\problem}& \makecell{Channel\\characteristics} \\ \hline
\end{tabular}
\end{table}

\begin{table*}[t]
\centering
\caption{Categorizing acoustic sensing enabled applications.}
\label{table:application comparison}
\begin{tabular}{|c|c|c|c|}
\hline
\textbf{Category}   &  \textbf{Key physical layer }   & \makecell{\textbf{Core processing techniques}}  & \textbf{Applications} \\ \hline
\makecell{Aerial acoustic \\ communication} & \makecell{Acoustic hardware,\\acoustic channel property} & Waveform design & Communication~\cite{AcousticCDMA,ChirpCommunication,AcousticNFC,AcousticOFDM} \\ \hline
\multirow{4}{*}{\makecell{Temporal feature\\based applications}} & \multirow{4}{*}{\makecell{Acoustic hardware,\\acoustic channel property}} &     
    \makecell{One-way sensing/\\ Phase-enabled accurate timing} & Ranging~\cite{BeepBeep,RFBeep,SwordFight} \\ \cline{3-4} 
    &  & 
    \makecell{Signal onset point detection,\\Phase-enabled accurate timing} & Acoustic radar~\cite{BatMapper,SAMS,DeepRange} \\ \cline{3-4} &  & 
    \makecell{Signal onset point detection,\\one-way/two-way sensing} & Localization~\cite{ALPS,ALPSPre,Guoguo,ARABIS,EchoTag,UPS,AALTS,Centaur,Liu} \\ \cline{3-4} 
    &  & 
    \multirow{3}{*}{Phase-enabled accurate timing} & 
        Device-based tracking~\cite{CAT,FollowMeDrone,AAMouse,MilliSonic} \\ \cline{4-4} 
        &  &  &  
        Device-free gesture tracking~\cite{FingerIO,LLAP,Strata} \\ \cline{4-4} 
        &  &  & 
        \makecell{Biometric sensing~\cite{BreathListener,SpiroSonic,NoiseRaspiration,RespTracker,SpiroSmart,CoughSensing} \\ \cite{StressSense,SleepApeaDetection,MusicalHeart,EmotionSense,SleepMonitoring}} \\ \hline
\multirow{2}{*}{\makecell{Channel characteristics\\enabled applications}} 
& \multirow{2}{*}{\makecell{Acoustic hardware,\\acoustic channel property,\\platform diversity}} & 
    \multirow{3}{*}{Over-the-air channel profiling} & 
        Gesture recognition~\cite{SoundWave,AudioGest,UltraGesture,LimitofGestureRecognition,VSkin,DepthAware} \\ \cline{4-4} 
        & & & 
        \makecell{Speaker authentication~\cite{BreathPrint}} \\ \cline{4-4}
        & & & 
        \makecell{Novel interactive control~\cite{Acoustruments}} \\ \cline{3-4} %\hline 
    %Over-the-air channel & \makecell{Gesture recognition~\cite{SoundWave,AudioGest,UltraGesture,LimitofGestureRecognition,VSkin,DepthAware},\\
    %speaker authentication~\cite{BreathPrint},\\
    %novel interactive control~\cite{Acoustruments}}  \\ \cline{3-4} 
    %& & 
    %Structure-borne channel & \makecell{Gesture recognition~\cite{SoundWave,AudioGest,UltraGesture,LimitofGestureRecognition,VSkin,DepthAware}}  \\ \cline{3-4}     
    &    &   
    \multirow{3}{*}{\makecell{Structure-borne channel profiling}} & 
        Keystroke detection~\cite{UbiTap} \\ \cline{4-4} 
        & & & 
        \makecell{Force detection~\cite{ForcePhone}} \\ \cline{4-4}
        & & & 
        \makecell{Touch recognition~\cite{TouchActive}} \\ \cline{4-4}
     %   & & & 
    %&  & 
    %PM \& DD & Keystroke detection~\cite{SnoopingKeystrokes,UbiK,KeystrokeAttack,UbiTap} \\ \cline{3-4} 
    %&  & 
    %PM or DD & Touch force detection~\cite{TouchForceSensing2,TouchForceSensing1,ForcePhone} \\ \cline{3-4} 
    %&  & 
    %DD & Authentication~\cite{BreathPrint,SpeakPrint}\\ \cline{3-4} 
    %&  & 
    %DD & Activity recognition~\cite{Auditeur,BodyBeat,DopEnc,CovertBand}\\ \cline{3-4} 
    %&  & 
    %DD & Interactive sensing~\cite{Auditeur,BodyBeat,DopEnc}\\ 
    \hline
\end{tabular}
\end{table*}

%The second important thing in channel modeling is to construct an effective model. There are often two ways, namely PM and DD, to develop the model and their respective advantages, disadvantages, and suitable applications are displayed in TABLE~\ref{tab:processing technique comparison}. 
Model-driven approach, formulating acoustic sensing problem as regression, can be effective and efficient, which however, require sophisticated signal processing designs and specific domain knowledge hence is often remarkable challenging. The key insight behind MD approaches is to mathematically quantify the relationship between CSI and certain application-specific states through regression, say mapping hand moving direction to specific Doppler frequency shift in a close-form equation. Actually, the hard part of this approach is to discovery the respective CSI and make it notable via signal processing techniques. Additionally, this approach may only be suitable when CSI is a scalar variable. Since these signal processing techniques are application-specific, we hence postpone their discussions in the next section. 

Since a close-form or explicit parametric model is often hard to craft, one often resort to DD approach. The DD approach, taking acoustic sensing as categorical problem, implicitly builds the inference model by directly mapping CSIs with respective application-specific states. 
This approach is more suitable to handle the cases when CSI is in the format of a vector, say frequency response, or matrix, say MFCC. 
The basic idea behind DD approach is fingerprinting. % and hence building acoustic sensing application via this approach is comparatively easier than PM-based ones. 
This approach first collects sufficient data (CSIs) under respective states in an offline manner and then utilize these data to train a model using machine learning techniques. The model is then used online to predict the corresponding state with respect to its input data. Such a processing pipeline may require less sophisticated domain knowledge hence is comparatively easier than MD approaches.  
Nevertheless, this approach requires massive data or constant calibrations due to device heterogeneity problem hence is labor intensive. Meanwhile, high computational cost and storage requirement are other drawbacks for this method. %Despite their differences, both approaches require intensive observations so as to discovery the relationship between CSI and human readable information. 

%DD approach often involves two phases: offline phase and online phase. During offline phase, sufficient data with labels is collected. If various platforms are included in the data collection process, platform diversity problem should be properly handled; otherwise system performance can be severely affected. To construct the model, machine learning techniques such as support vector machine~\cite{TouchActive,SnoopingKeystrokes} are popular choices. Deep learning techniques can be another powerful alternative. One can design appropriate deep learning models based on the dimensions of the input acoustic features. When the input feature is Short-time Fourier Transform (STFT), using Convolution Neural Network (CNN)~\cite{} would be more appropriate. If the acoustic features between consecutive inputs have temporal connections, Recurrent Neural Network~\cite{} or temporal convolution~\cite{} may be feasible methods. When the input is a one-dimensional vector, fully connected neural network is worth a try. \textcolor{green}{To train the model requires a loss and since it is application specific we thus leave the discussions in Section~\ref{}.} During the online phase, the model uses the learned experience to make predictions based on the new inputs.  

\subsubsection{Challenges for Estimating Channel Characteristics}
\begin{itemize}
    \item As having been outlined, mining the specific CSI to reflect particular application-specific state requires sophisticated signal processing techniques and domain knowledge hence is remarkable difficult. Therefore, the challenges is
    %\begin{displayquote}
        %\textit{\textbf{Challenge-X}: Discovering appropriate CSI representation.}
    %\end{displayquote}     
    \begin{challenge}\label{clg:appropriate CIR representation}
    Discovering appropriate CSI representation.
    \end{challenge}
    This may require strong background and intense observations. An optional choice is resorting to customized hardware that offers more flexible physical-layer reconfigurability, so as to gain more opportunity to enhance the acoustic features, thereby facilitating CSI extraction. More discussions on this can be found in Section~\ref{sec:future direction}.
    \item As we mentioned earlier, channel models, especially those using DD approaches, may be platform-dependent, so it requires constant calibrations or user configurations when deploying a particular acoustic sensing applications on heterogeneous platforms. As a result, it is particularly challenging in 
    %
    %\begin{displayquote}
    %    \textit{\textbf{Challenge-XI}: Building platform diversity agnostic acoustic sensing algorithms.}
    %\end{displayquote} 
    \begin{challenge}\label{clg:cross-platform models}
    Building cross-platform acoustic sensing channel models.
    \end{challenge}
    Potential solutions to this problem may be few shot learning techniques~\cite{FewshotLearning} or domain adaptation methods~\cite{DomainAdaptationSurvey} and more discussions on this can be found in Section~\ref{sec:future direction for channel characteristics}.
\end{itemize}

\section{Acoustic Sensing Applications}
\label{sec:solution or application} 
Now we turn to acoustic sensing applications driven by the mechanisms discussed in Section~\ref{sec:core acoustic mechanisms}. Existing applications fall into three categories according to their respective supporting mechanisms, namely aerial acoustic communication, applications leveraging temporal features, and solutions enabled by channel characterization. For each application, we review key enabling techniques and performance achieved. Potential future directions for each categories are also discussed. A taxonomy of acoustic sensing applications is given in Table~\ref{table:application comparison}.
%
%In this section, we discuss various acoustic-enabled applications. Based on the application scenarios, we classify existing research into three categories: \emph{context-aware application}, \emph{Human Computer Interaction (HCI)}, and \emph{aerial acoustic communication}. Different categories exploit acoustic signals in different manners. Context-aware applications, depending on contextual information such as range, location, etc., mostly are built on sound propagation time estimation.
%HCI systems infer and respond to user intentions by deciphering special acoustic signatures from physical activities. 
%Aerial acoustic communication utilizes acoustic signal as wireless medium to deliver content.
%
%These applications mostly adopt \emph{active} sensing where modulated acoustic waves are generated.
%A taxonomy of these applications is summarized in Table~\ref{table:overall comparison}. These applications mostly if not all are developed using commodity mobile devices including smartphone, wearables, and laptops, etc. The preferred operational frequency range typically lies in the inaudible range (above $18$ kHz~\cite{VSkin}) due to possible discomfort caused by audible noises. Sometimes, the audible frequency range may also be exploited to enhance performance for the high receiver gain. In the following section, we discuss the basic principles behind these applications, their performance, and shortcomings, etc.
\vspace{-.5ex}

\subsection{Aerial Acoustic Communication}
\label{sec:aerial acoustic communication via waveform manipulation}
Exploiting aerial acoustic channels or air-borne acoustic signals for communication, known as aerial acoustic communication, has attracted much attention recently. Aerial acoustic communication enables any device that has embedded microphones and speakers to communicate without extra hardware or complex network configuration. It can serve as an alternative to traditional RF-based device-to-device communication such as Bluetooth, Near Field Communication (NFC), and WiFi Direct. In a nutshell, aerial acoustic communication is realized by representing information as a function of different waveform configurations. As summarized in TABLE~\ref{table:communication}, different schemes mainly differ in the waveforms and bandwidth utilized, which result in different data rates, operational ranges and audibility.

% In this section, we first present existing aerial acoustic communication approaches targeting short-range and long-range applications, respectively. Then we discuss a new sensing paradigm as a future direction for acoustic communications.  
% via two categories, i.e., \textit{short-range high-throughput} and \textit{long-range low data rate} systems. We then introduce a new modulation technique that achieves simultaneously long-range and high-throughput, addressing the \textit{\textbf{Challenge-VI}}. Finally, we discuss a new sensing paradigm as a future direction for acoustic communications. 
%
\begin{table*}[b]
\vspace{-2ex}
\centering
\caption{Comparison of aerial acoustic communication systems.}
\label{table:communication}
\begin{tabular}{|c|c|c|c|c|c|}
\hline
\textbf{Work} & \textbf{Modulation mechanism} & \textbf{\makecell{Maximum \\ operating range\\(m)}} & \textbf{Bandwidth (kHz)} & \textbf{Audible or inaudible} & \textbf{Bit rate (bps)} \\  \hline
\cite{ToneBasedCommunication} & OFDM &  ${< 2}$  & ${0.735 - 4.41}$, ${18.4}$ & Audible or inaudible &  \makecell{5600\\1400} \\ \hline
Digital voice~\cite{MarrayFSK} & M-ary FSK &  ${< 2}$ & ${0 - 12}$  & Audible & ${2400}$  \\ \hline
\cite{AcousticMeshNetwork} & FHSS & ${20}$ & ${4.1 - 21}$   & Audible & ${20}$  \\ \hline
Dhwani~\cite{AcousticNFC} & OFDM & $<1$ & ${0 - 24}$  & Audible & ${2400}$  \\ \hline
~\cite{SignalEmbedding1} & OFDM & ${8}$  & ${6.4 - 8}$  & Inaudible & ${240}$  \\ \hline
~\cite{SignalEmbedding2} & \makecell{Phase modulation via \\ complex lapped transform} & ${< 2}$  & ${6.4 - 8}$   & Inaudible & ${600}$  \\ \hline
Dolphin~\cite{AcousticOFDM} & OFDM & ${8}$  & ${8 - 20}$  & Inaudible  & ${500}$  \\ \hline
~\cite{ChirpCommunication} & BOK & ${25}$  & ${19.5 - 22}$  & Inaudible &  ${16}$  \\ \hline
~\cite{AcousticCDMA} & QOK & \makecell{${2.7}$ m at \\ ${35}$ dBSPL } & ${18.5 - 19.5}$  & Inaudible & ${15}$  \\ \hline
\end{tabular}
\end{table*}

\label{sec:aerial acoustic communication}

\subsubsection{Short-range Aerial Acoustic Communication}
\label{sec:short range high throughput communication}
The authors in~\cite{ToneBasedCommunication} presented a communication system by manipulating the properties of pure tone signals. It leverages the presence or absence of tone signals to represent information (${100\%}$ Amplitude Shift Keying). This approach achieves a data rate of ${5.6}$~\!kbps with multiple audible tones. The data rate reduces to ${1.4}$~\!kbps when a single inaudible tone is used. A maximum communication range of ${2}$~\!m can be achieved under LOS conditions. Another work called Digital voice~\cite{MarrayFSK} modulates data bits in the audible band (under ${12}$~\!kHz) via M-ary FSK, reporting a data rate at tens to thousands of bits per seconds (bps). 
Dhwani \cite{AcousticNFC} is an acoustic-based NFC system. It employs OFDM modulations to encode messages and incorporate a  technique called JamSecure to prevent malicious attacking. Dhwani occupies a bandwidth of ${24}$~\!kHz and achieves a maximum data rate of ${2.4}$~\!kbps. The authors of~\cite{AcousticMeshNetwork} propose an acoustic-enabled mesh network built upon FHSS over 4.2KHz to 21KHz. The above work all utilize the audible band (normally below ${18}$~\!kHz). 
Transmitting information over audible bands can be disruptive and thus several inaudible (hidden) communication systems are developed. The authors in~\cite{SignalEmbedding1} propose to leverage the masking effect of the human hearing system to achieve inaudible acoustic communication, saliently addressing \textit{\textbf{Challenge}}-\ref{clg:audibility}. 
It employs OFDM modulations and achieves a data rate of ${240}$~\!bps. Similar work in~\cite{SignalEmbedding2} and~\cite{AcousticOFDM} attain data rates of ${600}$ and ${500}$~\!bps, respectively.

\subsubsection{Long-range Low Data Rate Communication}
\label{sec:long range low data rate communication}
Neither tone-based nor OFDM modulation techniques are robust to Doppler effects. The performance of these methods further deteriorates in multipath rich environments, making them inadequate for long range communications. In contrast, chirp spread spectrum (CSS) utilizes more interference-resilient chirp signals to encode information bits and thus achieves lower bit error rates and longer communication ranges.
A chirp binary orthogonal keying (BOK) modulation techniques is first presented in \cite{BOK1,BOK2}. It utilizes orthogonal up and down chirp signals for modulation. The work in~\cite{ChirpCommunication} adopts BOK and realizes a communication range up to ${25}$ m at a data rate of ${16}$ bps. Soonwon et al.~\cite{AcousticCDMA} improve upon BOK and develop a chirp quaternary orthogonal keying (QOK) modulation technique. QOK finds near-orthogonal chirps by an exhaustive search over a pre-defined solution space. With QOK, a Code Division Multiple Access (CDMA) system is built. The system achieves zero frame error rate even at a minimal sound pressure level of ${35}$ dB SPL when the transceivers are ${2.7}$ m away from each other. The authors of~\cite{AALTS} propose a pseudo-orthogonal CSS modulation where up- and down-chirps overlap in time and thus its transmission rate is doubled. CSS and its variants often achieve less bit error rates and long communication ranges compared with other modulation mechanism. However, the achievable data rate is less competitive as this method is inherently less bandwidth efficient.
%XXX: in the following statement, it is not clear to me whether the possible interference refers to the interference causes by using pseudo-orthogonal CSS or extra sources of interference 
%[CC]: Done
%Though possible interference is introduced, the reported performance does not degrade at all. 
%The advantages in bit error rates and communication ranges of CSS and its extensions come at the cost of low data rates. 

%Current aerial acoustic communication systems cannot strike a good balance between long communication distance and high data rate. Despite the above dilemma, most solutions tailored for \textcolor{red}{acoustic NFC} are already mature for commercial adoption. The aerial acoustic communication technology, both cost-effective and power-efficient, would certain enjoy plausible adoption for the booming internet-of-things networking.
%
\begin{table*}[b]
\centering
\caption{Comparison of different ranging techniques.}
\label{tab:ranging techniques comparison}
\begin{tabular}{|l|l|l|l|l|}
\hline
\textbf{Ranging solution} & \textbf{Key processing techniques} & \textbf{Waveform design} & \textbf{Occupied bandwidth (kHz)} & \textbf{Ranging error} \\ \hline
BeepBeep~\cite{BeepBeep} & \makecell{Two-way sensing} & Chirp signal & $2-6$  & An average around 1 to 2~\!cm within 10~\!m\\ \hline
RFBeep~\cite{RFBeep} & \makecell{One-way sensing } & Tone & NA &  Around 30~\!cm median error within 16~\!m\\ \hline
SwordFight~\cite{SwordFight} & \makecell{Two-way sensing } & Tone & $0-11.025$  or $0-16$  & A media of 2~\!cm within 2~\!m distance\\ \hline
BatMapper~\cite{BatMapper} & \makecell{Finer temporal feature} & Chirp signal & $8-16$  and $8-10$  & Up to 2~\!cm within 4~\!m distance\\ \hline
SAMS~\cite{SAMS} & \makecell{Finer temporal feature} & Chirp signal & $11-21$  & \makecell{A median of 30~\!cm \\and a $90$-percentile of $100$ cm within 5~\!m}\\ \hline
DeepRange~\cite{DeepRange} & \makecell{Finer temporal feature} & Chirp signal & $18-22$  & \makecell{A median of 1~\!cm within 4~\!m\\}\\ \hline
\end{tabular}
\end{table*}

\subsubsection{Long-range High-speed Communication via Loose Orthogonal Modulation}
\label{sec:long range high speed communication via loose orthogonal modulation}
To achieve long-range communication at relatively high data rates  (addressing \textit{\textbf{Challenge}}-\ref{clg:comm_range}), 
a loose orthogonal modulation approach is proposed in~\cite{HRCSS}. %It can be deemed as a high-speed version of traditional CSS modulation techniques. 
Its basic idea is to overlap multiple chirp carriers in a single time slot to boost throughput. Though inter-carrier-interference is introduced, its adverse effect can be mitigated, say, by rate adaptation. The reported maximum throughput is 1~\!kbps, over $60\times$ of that of existing CSS-based approaches. Even at a distance of 20~\!m, a data rate of 125~\!bps can be reached. 

\subsubsection{Research Opportunities} 
\label{sec:future direction for waveform design}
%
%\textcolor{red}{the future direction should have connection with challenges}
%
Aerial acoustic communication is gaining attraction as an alternative for device discovery and communication. In fact, audio interfaces has become a part of Google's nearby peer-to-peer communication APIs~\cite{Nearby}. 
However, low data dates continue to hamper its wide adoption. For instance, in-vehicle networking demands at least 10~\!kbps data rate~\cite{InVehicleNetwork}, 
%XXX: what is in-vehicle networking? for entertainment? control?? any citation?
%[CC]: Done
far exceeding the maximum throughput attainable by existing aerial acoustic communication systems. Data rates of aerial acoustic communication can be further increased by incorporating more complex techniques such as rake receivers, multiple-input-multiple-output, and advanced coding mechanisms. Another opportunity arises from the exploitation of SS channels as discussed in Section~\ref{sec:acoustic channel property}. Most existing works are restricted to AS for communication while SS is rarely considered with the only exceptions are \cite{Ripple} and \cite{RippleII}. In~\cite{Ripple}, the authors  modulate the vibration motors available in  mobile phones, and decoding information through accelerometers. A maximum of 200~\!bps data rate is achieved. In their follow-up work~\cite{RippleII}, the throughput is boosted to 30~\!kbps by replacing the receiver with a high-sensitivity microphone and the adoption of high-order OFDM modulations. 
We envision that more efficient modulation techniques for SS channels can be devised. Furthermore, combining AS and SS in a single system poses new challenges and exciting research venues in acoustic communication.  
%\subsection{Temporal Features based Applications}
\subsection{Applications Leveraging Temporal Features}
\label{sec:temporal feature based applications}
In this section, we present applications leveraging temporal features including acoustic ranging, localization, device-based tracking, device-free gesture tracking, and respiration sensing. For each application, we discuss state-of-the-art solution approaches and how some the challenges outlined in previous sections are addressed. 

\subsubsection{Ranging}
\label{sec:ranging}
Range (the relative distance between two devices) provides a useful contextual information and can be used for distance and size measurements, network management~\cite{NetworkManagement}, and content sharing~\cite{Photoware1,Photoware2}. Leveraging acoustic signals for ranging is an economical and convenient alternative to traditional measurement tools. Ranges are computed by multiplying time of flight derived from techniques in Section~\ref{sec:temporal features} with the propagation speed of acoustic signals, generally assumed to be constant and known {\it a priori}. \textit{Ranging errors} are defined as the Root Mean Square (RMS) of the differences between estimated results and the ground truth, while \textit{ranging accuracy} is inversely proportional to the RMS error. The ranging performance heavily relies on techniques at the processing layer, in particular, robust onset detection. A comparative study is shown in TABLE~\ref{tab:ranging techniques comparison}. 

%In this section, we talk about a ranging technique
%that acquires the relative distances between pair-wised devices by measuring the time it takes for acoustic signals to travel from one device to another. Since the propagation speed of sound is known, multiplying the speed by the time can thus get range information. Often, the performance of this technique depends on the sophisticated signal designs at the physical layer and robust preamble detection approaches at the processing layer. A comparative study is shown in TABLE~\ref{tab:ranging techniques comparison}. 

BeepBeep~\cite{BeepBeep} is a pioneer work that uses acoustic signals for precise ranging on commodity mobile devices based on ToA via a two-way sensing method. It cleverly circumvents unknown system delays (\textit{\textbf{Challenge}}-\ref{clg:system_delay}) and clock synchronization by directly retrieving timestamps from acoustic samples themselves. To mitigate multi-path effects aroused from \textit{\textbf{Challenge}}-\ref{clg:robust onset detection}, the authors locate the earliest ``sharp" peak after cross-correlation. 
BeepBeep reports centimeter-level (2 to 6~\!cm) ranging errors. However, the performance of BeepBeep can be degraded by presence of strong NLOS paths.
%XXX: I remove irregular & uncertain system delay since BeepBeep does not suffer from the problem
%[CC]: ok
The authors in~\cite{RFBeep} sidestep uncertain and variable system delays in acoustic playback and recording hence addressing \textit{\textbf{Challenge}}-\ref{clg:system_delay} by a kernel space implementation, and build a stand-alone application called RFBeep.
RFBeep~\cite{RFBeep} performs ranging with ToA using one-way sensing.  %\textcolor{red}{The key idea for RFBeep is that the flight time of radio signals is negligible in the maximum reachable distance for power-limited acoustic signal. Thus when triggered simultaneously, radio signals is used for synchronization.} As a result, the range information can be acquired via estimating the flight time of acoustic signal in one-way sensing manner. 
It reports decimeter-level (up to 50~\!cm) ranging errors within 16~\!m. However, the reliance on kernel modification prohibits its wide-scale adoption. 
SwordFight~\cite{SwordFight} is another ranging solution that improves upon BeepBeep in responsiveness, accuracy, and robustness. It works at the same way as BeepBeep but differs in transmitted waveforms and onset detection strategies. It reports a median ranging accuracy of ${2}$~\!cm with ${12}$~\!Hz fresh rate in noisy environments. 
%XXX: same comment regarding system delay
%However, SwordFight does not resolve system delay. %In addition to the above mentioned problems, the Sampling Frequency Offset (SFO) caused by hardware heterogeneity is a common issue that affects the ranging accuracy. Also, the limited operational range due to the adoption of severely attenuated inaudible range and annoying disturbance caused by using audible range are another two major concerns.

%To summarize, the commonly used asynchronous ranging techniques suffer from uncertainty system delay that might significantly affect the system performance. Another problem is SFO caused by hardware heterogeneity, which however, does not introduce noticeable ranging errors. Often, this category of ranging techniques need to detect a particular reference signal, known as preamble detection detailed in Section~\ref{sec:processing layer}, and use the time it occurs to determine timestamps for ranging. Therefore, to achieve high ranging results, a dedicated signal design at the physical layer and reliable preamble detection strategy at the processing layer are desirable.

%\subsection{Channel Response enabled Applications}
%Context-aware applications, built on contextual information such as range or location can provide better user experience in many applications such as health, entertainment, etc. The context-aware applications, mainly built on parametric modeling, can be further grouped into four categories: ranging, acoustic radar, device-based tracking, and localization.

\begin{table*}[b]
\centering
\caption{Comparison of acoustic-enabled localization systems.}
\label{table:localization comparison}
\begin{tabular}{|c|c|c|c|c|c|c|c|}
\hline
     \textbf{Category}             & \textbf{Work} & \textbf{\makecell{Waveform\\design}} & \textbf{\makecell{Additional\\signal}} &  \textbf{\makecell{Key processing\\ technique}} & \textbf{\makecell{Synchronous or\\ asynchronous}} &\textbf{ \makecell{Concurrent\\localization}} & \textbf{\makecell{Localization\\error}} \\ \hline
\multirow{4}{*}{\makecell{Infrastructure\\-based}} & Guoguo~\cite{Guoguo} & \makecell{Gaussian\\duplet pulse} & NA & ToA (one-way) & Synchronous & Supported & Centimeter-level \\ \cline{2-8}
                  & ~\cite{ALPSPre} & Chirp & NA & TDoA (one-way) & Synchronous & Supported & Centimeter-level \\ \cline{2-8}
                  & ALPS~\cite{ALPS} & Chirp & NA &TDoA (one-way) & Synchronous & Supported & Decimeter-level \\ \cline{2-8}
                  & UPS+~\cite{UPS} & \makecell{Ultrasonic\\chirp, tone} & NA & ToA (one-way) & Synchronous & Supported & Centimeter-level \\ \cline{2-8}
                  & ARABIS~\cite{ARABIS} & Chirp & NA & TDoA (two-way) & Asynchronous & Supported & Centimeter-level \\ \cline{2-8}
                  & AALTS~\cite{AALTS} & \makecell{Chirp,\\pure tone} & NA & TDoA (two-way) & Asynchronous & Supported & Decimeter-level \\ \hline
\multirow{3}{*}{\makecell{Infrastructure\\-free}} & ~\cite{Liu} & Chirp & WiFi & ToA (two-way) & Asynchronous & Not supported & Meter-level \\ \cline{2-8}
                  & Centaur~\cite{Centaur} & Chirp & WiFi & TDoA (two-way) & Asynchronous & Supported & Meter-level \\ \cline{2-8}
                  & EchoTag~\cite{EchoTag} & Chirp  & NA & Data-driven & N/A & Not Supported & Centimeter-level \\ \hline
\end{tabular}
\end{table*}

\subsubsection{Acoustic Radar}
\label{sec:acoustic radar}
Similar to RF radars, acoustic radars work by radiating acoustic signals and estimating the round trip time of reflected echoes. It can be used in obstacle avoidance and mapping. Since acoustic echoes attenuate sharply with the increase of distance $d$ (e.g., in free space, proportional to $\frac{1}{d^4}$~\cite{Strata}), the operational range of acoustic radars is often limited.  
%XXX: why d^4?
%[CC]: just cite it from other papers. 
Multipath effects also pose additional challenges as it is often difficult to disentangle desired echoes bouncing off targeted objects from those experiences multiple reflections or reflected off undesired objects (\textit{\textbf{Challenge}}-\ref{clg:robust onset detection}). The evaluation metrics for acoustic radars are the same with the one used in ranging.
 
%\begin{figure*}
%    \centering
%    \includegraphics[width = 5.85in]{Figs/microphone_layouts_for_survey.pdf}
%    \caption{Typical microphone and speaker layouts for smartphones.}
%    \label{fig:sensor layout}
%\end{figure*}

%\section{\textcolor{green}{Solutions/Applications}}
%\label{sec:application layer} 

In  \cite{SonarRanging}, the authors proposed a sonar sensor for depth sensing on smart phones utilizing the phone's microphone and rear speaker. The reported ranging errors up to  ${12}$~\!cm within ${40}$~\!m distances. Further improvement to ranging accuracy using deep learning techniques is proposed in~\cite{DeepRange}. In this work,
Synthesized data is used to train a neural network to handle platform diversity, multipath effects, and background interference, addressing \textit{\textbf{Challenge}}-\ref{clg:calibration} and \textit{\textbf{Challenge}}-\ref{clg:robust onset detection}
%\textit{\textbf{Challenge-II}, %\textit{\textbf{Challenge-IV}, %\textit{\textbf{Challenge-XI}}, and %\textit{\textbf{Challenge-VII}}}}
simultaneously. The resulting range error can be as low as 1~\!cm at a distance up to 4~\!m and is agnostic to various background noises, heterogeneous devices, etc. 
BatMapper \cite{BatMapper} is among the first work that demonstrates the feasibility of commodity mobile devices to act as acoustic radars for indoor floor map construction. It exploits speaker-microphone distance constrains to construct a probabilistic model to extract echoes bouncing off surrounding objects, which allows it to mitigate multipath effects (\textit{\textbf{Challenge}}-\ref{clg:robust onset detection}) for robust onset detection and thereby achieve accurate range estimation. BatMapper reports ${1\!-\!2}$~\!cm estimation errors with ranges up to ${4}$~\!m. With the assistance of on-board Inertial Measurement Unit (IMU) including a gyroscope and an accelerometer, its  $80$-percentile errors are less than $30$~\!cm in geometric floor reconstruction. 
SAMS~\cite{SAMS} is another acoustic radar solution that improves upon BatMapper. Different from the correlation method adopted in BatMapper, SAMS utilizes chirp mixing, a finer temporal feature extraction method, which can circumvent insufficient sampling rates for better temporal resolution. SAMS reports a median error of $30$~\!cm and a $90$-percentile error of $1$~\!m. 

Compared to ranging between acoustic devices, acoustic radars rely on weak reflected signals from target objects for ranging measurements. Therefore, it is much more challenging to design appropriate signal processing techniques and robust acoustic waveforms for acoustic radars. A technical comparison among different ranging approaches is given in TABLE~\ref{tab:ranging techniques comparison}. 
%Both ranging and acoustic radar obtain only one-dimensional distance information. To obtain position information in 2D or 3D coordination, one needs to resort to localization.  

\subsubsection{Localization}
\label{sec:localization}
Localization is a key enabler for Location Based Service (LBS). Despite tremendous research efforts on indoor localization, many existing solutions either require expensive dedicated infrastructures~\cite{ArrayTrack,WiTrack,iLocScan} or rely on cumbersome device-dependent kernel hacking~\cite{Chronous,Ubicare,SpotFi,M3}, prohibiting their practical deployment. 
Among existing cutting-edge indoor localization approaches, acoustic-based systems attract much interests since they can achieve sub-meter level localization accuracy with relatively low infrastructure costs and deployment efforts. The underlying techniques for these localization systems are the timing measurement methods discussed in Section~\ref{sec:basic timing measurements} and the major challenge is \textit{\textbf{Challenge}}-\ref{clg:robust onset detection}. The evaluation metric for localization systems is RMS error. 
%include the median, average, or Cumulative Distributed Function (CDF) of RMS errors.
In this section, we present existing works on acoustic-enabled localization solutions in two categories, namely, \emph{infrastructure-based} and \emph{infrastructure-free}. A comparison of related work is summarized in TABLE~\ref{table:localization comparison}.

\textbf{Infrastructure-based} schemes typically deploy low-cost and power-efficient distributed acoustic anchors in target areas. The locations of these anchors are determined in advance. Apart from acoustic transceivers, each anchor may be equipped with wireless modules to communicate among themselves or with a remote server. The remote server can coordinate the transmission schedule among the anchors either in synchronous or asynchronous manner. When the transmitted acoustic signals are detected by either a target or other anchors, the associated timestamps (ToA or TDoA) are obtained. Finally, the location of a target is determined using trialeration or more sophisticated optimization methods. 
In the subsequent discussion, we first review synchronous schemes, and then asynchronous approaches.

In ~\cite{Guoguo}, Liu et al. developed a centimeter-level localization system named Guoguo. The anchors in this system are synchronized by Zigbee and are scheduled to transmit orthogonal codes, which are used by targets to perform ToA estimation via one-way sensing. Multilateration is then used to locate the targets.
A speaker-only localization system was proposed by Lazik and Rowe in~\cite{ALPSPre}. In their approach, distributed speakers are connected to different synchronized channels of an advanced audio device that transmits chirp signals for localization. A target locates itself locally by performing one-way TDoA estimation. According to \cite{ALPSPre}, its ${95}$-percentile localization accuracy is within ${10}$~\!cm. ALPS~\cite{ALPS} improves upon the work~\cite{ALPSPre} in deployment efforts. In ALPS, anchors are synchronized via Bluetooth, and each anchor is equipped with one microphone and one speaker. The locations of the anchors are efficiently obtained through acoustic-assisted simultaneously localization and mapping. ALPS reports average errors of ${30}$~\!cm and ${16.1}$~\!cm in locating targets and anchors. Recently, a ultrasonic localization system called UPS+ is presented in~\cite{UPS}. UPS+ leverages the non-linearity of receiver microphones (details are presented in Section~\ref{sec:future direction for channel characteristics}) to enable ultrasonic beacons to locate smart devices without ultra-sonic sensors. Consequently, the audibility problem, induced either by \textit{\textbf{Challenge}}-\ref{clg:audibility} or \textit{\textbf{Challenge}}-\ref{clg:timing estimation trade-off}, is eliminated. UPS+ achieves centimeter-level accuracy in localization. 

\begin{table*}[b]
\caption{Comparison of device-free gesture tracking systems.}
\label{table:device-free gesture tracking comparison}
\begin{center}
    \begin{tabular}{ | c | c | c | c | c | c |}
    \hline
    \textbf{System} & \textbf{Waveform Design} & \textbf{\makecell{Occupied bandwidth\\(kHz)}} & \textbf{\makecell{Tracking latency (ms)}} & \textbf{\makecell{Operation range\\(m)}}& \textbf{\makecell{Performance}} \\ \hline
    FingerIO~\cite{FingerIO} & OFDM modulated signal  & ${18 - 20}$ & ${5.92}$  &  ${< 0.5}$ & \makecell{${8}$~\!mm (2D) average\\tracking error}\\ \hline
    LLAP~\cite{LLAP} & Multiple pure tones & ${17 - 23}$  & ${\le 15}$   & ${0.5}$ & \makecell{${3.5}$~\!mm (${1}$D) and ${4.57}$~\!mm (${2}$D)\\average tracking error} \\ \hline
    Strata~\cite{Strata} & GSM sequence   & ${18 - 22}$  & ${12.5}$  & ${0.5}$  & ${3}$~\!mm tracking error \\ \hline
    \cite{MotionTracking} & Chirp signal   & ${18 - 20}$  & ${40}$  & ${4.5}$ & ${1.2}$ to ${3.7}$~\!cm within 4.5~\!m\\
    \hline
    CovertBand~\cite{CovertBand} & OFDM   & ${18 - 20}$  & ${4.2}$  & ${6}$ & \makecell{A median of ${18}$~\!cm\\tracking error} \\
    \hline
    \end{tabular}
\end{center}
\end{table*}

The localization accuracy of the aforementioned work is highly dependent on clock synchronization accuracy, which depends on network latency, non-negligible in a large-scale network. In contrast, asynchronous approaches can overcome such shortcomings.
ARABIS \cite{ARABIS} is an asynchronous acoustic localization system that utilizes two-way ranging~\cite{BeepBeep} to avoid the need for synchronization. In ARABIS, anchors transmit acoustic beacons periodically following a coarse time-division-multiple-access schedule. Targets, as well as anchors, overhear the transmissions and record the corresponding timestamps. These timestamps can be used to estimate TDoA information in locating a target. ARABIS reports a ${95}$-percentile localization error of ${7.4}$~\!cm. AALTS~\cite{AALTS} improves upon ARABIS by a more robust onset detection approach to handle the near-far problem, hardware heterogeneity, and multipath effects. To handle these challenges (from \textit{\textbf{Challenge}}-\ref{clg:robust onset detection}), it normalizes the current correlation value by the mean of a number of its preceding samples. Additionally, a pseudo orthogonal chirp spread spectrum modulation technique is proposed, which effectively doubles the transmission rate. AALTS achieves $90$-percentile tracking errors of $0.49$~\!m for mobile targets and a median of $0.12$~\!m for stationary ones with only four anchor nodes. %\textcolor{blue}{talks about some layout issues here.}

\textbf{Infrastructure-free} localization systems do not require the deployment of custom-built infrastructure devices in target areas. However, they tend to achieve less competitive localization accuracy compared with infrastructure-based solutions.

In~\cite{Liu}, Liu et al. built a localization system utilizing acoustic and WiFi signals. It first estimates pair-wise distances within a device group via acoustic ranging~\cite{BeepBeep}, forming a spatial constraints. 
Each device in the group also uses WiFi fingerprints to impose another location constraints. By combining the two, target locations can be determined.
This scheme achieves an ${80}$-percentile localization error of ${1}$~\!m. Since the computation of  spatial constraints requires multiple pair-wise acoustic ranging  measurements and is time consuming, the application of such an approach is limited to  static target localization. 
Centaur~\cite{Centaur}, similar to \cite{Liu}, is a joint optimization framework utilizing acoustic and WiFi signals, and reports meter-level localization accuracy. The authors propose a novel multipath mitigation algorithm to address \textit{\textbf{Challenge}}-\ref{clg:robust onset detection}, and achieve robust onset detection. The key idea is to inspect signal changes in cross-correlation as opposed to absolute magnitudes considered in existing methods. 
EchoTag~\cite{EchoTag} is an acoustic  fingerprinting localization system that can detect minor location changes. It associates different acoustic profiles with different positions, known as tags, to train a classification model. This model is then used for online tag detection, enabling context-aware applications. 
EchoTag reports an accuracy of ${98\%}$ in distinguishing ${11}$ tags at ${1}$~\!cm resolution. However, EchoTag is sensitive to environmental dynamics and will suffer from degraded performance in absence of new data collections. As a matter of fact, applications that are based on a fingerprinting strategy are subject to problems aroused from \textit{\textbf{Challenge}}-\ref{clg:calibration} and \textit{\textbf{Challenge}}-\ref{clg:cross-platform models}, limiting their practical adoption. 
 
%XXX: I have to remove this because the two works are for acoustic event localization. if you include them, this will open a whole can of worms on sound localization using microphone array and distributed acoustic sensors
% [CC]: But in conventional microphone array based localization, they can only obtain AoA information, not its coordination. I can add some comments after I discuss these works. 

Although infrastructure-free solutions incur less hardware costs, they require labor-intensive site survey to obtain location-dependent signal profiles, making them sensitive to environment changes. Infrastructure-based approaches deliver a satisfactory localization accuracy at the cost of extra hardware. But the complexity in deploying multiple acoustic anchors and the synchronization requirements are still not economical and lightweight enough for practical deployment. To this end, the authors of~\cite{PACE} propose a single beacon-enabled passive localization system that can also identify a target. They discover that a footstep contains separable structure-borne and air-borne components. The former contains range information and the latter provides angle-of-arrival (AoA) along with identity signatures. Consequently, by placing a single acoustic array in the place of interest, a target can be simultaneously tracked and identified. Additionally, the domain adversarial training technique is employed in this proposal so as to enhance the generalizability of the system, easing the efforts in calibration and thus addressing \textit{\textbf{Challenge}}-\ref{clg:cross-platform models}. The reported median localization accuracy can reach 30~\!cm, which is highly enough given that a foot has a similar size. Another single acoustic anchors based proposal that leverage the geometry constraints shaped by LoS and NLoS acoustics can be found in~\cite{Voloc} that reports 0.44~\!m localization accuracy across different environments. It is worth to mention that the aforementioned microphone array enabled localization techniques can obtain specific coordinates of a target rather than conventional source localization that can only obtain AoA information. 

%Infrastructure-free solutions can be supplementary to the infrastructure-based systems when anchor nodes cease to operate and do not have sufficient coverage. However, infrastructure-free approaches require tedious site survey to obtain location-dependent signal profiles, making them quite sensitive to environment changes.
%Most infrastructure-free and infrastructure-based acoustic localization techniques are designed for stationary targets.  
%To facilitate tracking, especially high accuracy tracking, different solutions are needed.

Compared to localization solutions utilizing RF signals~\cite{MeWiFiSensing}, visible light~\cite{VLC1}, or IMU data~\cite{IMU1}, acoustic-enabled localization techniques strike good trade-offs between costs and accuracy. The acoustic diffraction property makes it feasible to locate targets in presence of small-scale random blockages. This puts less restriction on deployment. However, due to limited transmission ranges, more anchor nodes are needed in infrastructure-based localization systems.   
%We envision that further localization techniques would not only deliver highly accurate localization results but also has lower system complexity. 
%The major bottleneck is the needs for line-of-sight paths. % and relatively small coverage areas.  
%

\begin{table*}[b]
\centering
\caption{Comparison between different biometric sensing systems.}
\label{tab:biometric sensing comparison}
\begin{tabular}{|c|c|c|c|c|}
\hline
\textbf{Work} & \textbf{Waveform design}  & \textbf{Bandwidth} & \textbf{Key techniques} & \textbf{Performance}\\ \hline
\cite{SleepApeaDetection} & FMCW & $18-20$~\!kHz & FFT &  \makecell{Less than 0.11~\!bpm\\within 1m} \\ \hline
BreathJunior~\cite{NoiseRaspiration} & FMCW \& white noise & 24~\!kHz & \makecell{Phase-based accurate timing,\\beamforming} & \makecell{0.4~\!bpm at 40~\!cm,\\3~\!bpm at 60~\!cm}\\ \hline
RespTracker~\cite{RespTracker} & ZC & 2~\!kHz & \makecell{Phase-based accurate timing} & \makecell{Less than 1~\!bpm at 3~\!m,\\0.8~\!bpm for moving targets}\\ \hline
\cite{SleepMonitoring} & NA & NA & \makecell{Envelop detection} & \makecell{Less than 0.05~\!bpm\\(device close to user)}\\ \hline
BreathListener~\cite{BreathListener} & tone at 20~\!kHz  & NA & \makecell{Energy spectrum density,\\ensenmble empirical mode decomposition,\\generative adversarial network} & \makecell{0.11~\!bpm\\in driving environment}\\ \hline
SpiroSonic~\cite{SpiroSonic} & Multiple tones  & $17-24$~\!kHz & \makecell{Phase-based accurate timing,\\neural network regression} & \makecell{5\%-10\% error in\\lung function monitoring}\\ \hline
\end{tabular}
\end{table*}

\subsubsection{Tracking}
\label{sec:tracking}

High-accuracy object tracking is important in many applications such as automated surveillance, traffic monitoring, and Augmented/Mixed Reality~\cite{ObjectTracking}.
Tracking is a well-investigated topic in computer vision (CV)~\cite{ObjectTracking,ObjectTracking1,ObjectTracking2}. However, CV techniques impose substantial computation costs and do not work well under poor light conditions. Acoustic tracking systems can overcome these limitations. Depending on whether the targeted object can emit acoustic signals or not, existing solutions can be divided into device-based tracking and device-free tracking. 

{\bf Device-based acoustic tracking} aims to track acoustic emitting devices in motion. 
%\textcolor{green}{DopEnc \cite{DopEnc} is an automatic encounter profiling system. It enables users to record conversation events and interaction contexts with other people automatically. The underlying techniques of DopEnc are Doppler frequency estimation and self-voice recognition. DopEnc achieves an accuracy of ${6.9\%}$ false positive and ${9.7\%}$ false negative rates in real-world usage. }
%BodyBeat~\cite{BodyBeat} is a mobile sensing system that aims to recognize non-speech body sounds such as food intake, laughter, and breath. This approach requires users to wear a dedicated device around the neck to capture audio samples. It builds a classification model that employs ${30}$ acoustic features for identification.
AAMouse~\cite{AAMouse} utilizes multiple carriers to estimate the Doppler speed of a target and integrates the speed over time for tracking. Though the reported tracking performance is at centimeter-level, tracking errors can accumulate over time, making it unsuitable for long-term tracking. CAT~\cite{CAT} improves upon AAMouse by adopting chirp mixing and boosts the tracking accuracy to a sub-centimeter level. However, due to the use of one-way sensing, it is sensitive to irregular SFOs~\cite{SFO2,SFO3,SFO1} that cannot be easily compensated. Under the assumption of linear drift, CAT performs re-calibration when integration errors become intolerable hence allow the system to work sufficiently long. 
%Therefore, CAT needs to perform calibrations from time to time. 
Another application of chirp mixing for high-accuracy tracking is presented in~\cite{FollowMeDrone} where a drone follows a person with a safe range in challenging indoor environments. In this work, the authors introduce several advanced signal processing modules, in particular, MUlitple SIgnal Classification algorithm (MUSIC) to resolve multipath effects (aroused from \textit{\textbf{Challenge}}-\ref{clg:robust onset detection}) and enhance system robustness. Furthermore, a reciprocal filter is introduced to address \textit{\textbf{Challenge}}-\ref{clg:freq_sel}, compensating the frequency selectivity problem and thereby further enhancing the system stability.
In Backdoor~\cite{BackDoor}, a more precise compensation technique for frequency selectivity is proposed. First adopted in wireless communication, it equalizes channel effects by measuring channel state information  using probe signals. Backdoor transmits acoustic signal in 40~\!kHz and takes advantages of non-linear diaphragm of power-amplifier so that sounds in the range of 20~\!kHz can be recorded. Thus, it does not suffer from the audibility problem ({\it{\bf Challenge}}-\ref{clg:audibility} or \textit{\textbf{Challenge}}-\ref{clg:bw_tradeoff}). However, this approach requires customized hardware devices, making its adoption more difficult. 
MilliSonic~\cite{MilliSonic} achieves sub-millimeter 1D tracking accuracy in the presence of multipath using a single beacon with a small 4-microphone array. The high precision is achieved by leveraging phase information after chirp mixing. 
%The underlying premise is that phase enables tracking metrics in continuous format while previous approaches are deemed as a discrete sampling from the metrics. More discussions on this can be found in Section~\ref{sec:temporal feature via phase}.

%Device-free tracking via acoustic sensing is less common due to limited signal power and severe path loss attenuation. 

%As covertBand uses parametric modeling, it limits to recognize only rhythmic motions.

%\textcolor{red}{Acoustic-enabled tracking techniques require LoS path in order to achieve highly accurate results and the performance would degrade sharply if with blockages.}
%Other challenges include limited processing capabilities and battery power on mobile devices. Sophisticated signal processing pipelines required for acoustic tracking can incur substantial latency and energy consumption.

%\begin{figure}
%    \centering
%    \includegraphics[width = 0.455\textwidth]{Figs/interactive_sensing5.pdf}
 
%    \caption{\textcolor{green}{HCI applications with acoustic sensing. (the figure should be reploted)} }
%    \label{fig:interactive sensing}
%\end{figure}

%\subsubsection{Gesture Tracking}
%\label{sec:gesture tracking}
{\bf Device-free acoustic tracking} tracks moving objects in an environment by reflected acoustic signals. Due to significant attenuation of reflected signals, high precision device-free acoustic tracking is often limited to short ranges. Next, we use finger, body posture tracking and respiration sensing as driving applications to discuss techniques in this category. 

%extends interaction space beyond the physical boundary of small mobile devices and effectively uses the nearby 3D space for interaction. This kind of technology is particularly useful for small portable devices such as wearables. Gesture recognition systems are essentially based on short-range acoustic radars. The key processing techniques behind these applications are that moving fingers can generate identifiable acoustic echoes, whose flight time can be accurately estimated via phase-enabled accurate timing described in Section~\ref{sec:phase-enabled accurate timing}. In this section, we present various gesture tracking systems based on different waveform designs. 

FingerIO turns a mobile phone or a smartwatch into an active sonar that is capable of tracking moving fingers for Around Device Interaction (ADI). Such a technology can extend the physical interaction boundaries hence is particular useful for small wearables. FingerIO achieves a median accuracy of ${8}$ mm~\cite{FingerIO} in 5.92~\!ms. 
It utilizes OFDM modulated signals to estimate the CSI between a hand and a smartphone periodically. In each estimation cycle, the CSI is acquired through cross-correlation. Since only moving fingers can dynamically affect the channel between the speaker and microphone, their movements can be tracked by comparing consecutive channel frames. Static multipath reverberations remain the same across frames and can thus be removed. The proposed technique successfully addresses \textit{\textbf{Challenge}}-\ref{clg:robust onset detection} by extracting finger-movement-only CSI profiles, and inspired several follow-up work.

A multi-tone device-free gesture tracking system, LLAP, was proposed in~\cite{LLAP}. It leverages coherent detection to extract the phases of acoustic echoes for finger localization and tracking. In LLAP, a mobile device actively transmits multiple tone carriers and decomposes finger-generated echoes via Empirical Mode Decomposition (EMD) for later processing. It further uses the phase divergence of multiple carriers to coarsely locate the start position of a finger and track its displacement via phase shifts.
LLAP reports a tracking accuracy of ${3.5}$ mm for ${1}$D hand movement and ${4.57}$ mm for ${2}$D drawing with less than ${15}$ ms latency. However, both FingerIO and LLAP are sensitive to nearby interference. To address nearby interference, another work named Strata was proposed in~\cite{Strata}. 
It also uses a coherent detector but applies a GSM training sequence modulated by Binary Phase Shift Keying (BPSK). 
Evaluation results demonstrate that it outperforms FingerIO and LLAP in all cases with an average tracking accuracy at 3~\!mm. 

The aforementioned work considers micro-finger gesture tracking. For macro-body parts such as hand or the whole body, one often utilizes more powerful speakers to increase SNR so that the sensing range is larger. We hereby present tracking technologies on macro-body posture. 
The work in~\cite{MotionTracking} shows that it is feasible to achieve room-level hand motion tracking with a customized platform. It employs an acoustic radar along with many advanced processing techniques including MIMO beamforming and deep learning for signal quality enhancement. The proposed system achieves $1.2$-$3.7$~\!cm tracking errors within $4.5$~\!m range and supports multi-user tracking. CovertBand~\cite{CovertBand} is an active sensing system for passive multiple object tracking. It builds on an active sonar with an enhanced speaker and uses the same parametric models as FingerIO~\cite{FingerIO} to track human body posture. CovertBand reports a median of $18$~cm in tracking mobile targets. For static objects, it can achieve an accuracy of $8$~cm with a distance up to $8$~m in LOS conditions. 
A comparison of these tracking schemes is given in TABLE~\ref{table:device-free gesture tracking comparison}.

The last category of applications concern continuous monitoring of human's respiration rates and breathing patterns. With device-free acoustic tracking, one can estimate chest displacements caused by respiration over time of target subjects. A comparison of these techniques is shown in TABLE~\ref{tab:biometric sensing comparison}.

In~\cite{SleepApeaDetection}, a portable life sign detection system based on commodity smartphones was presented. It uses a smartphone as an active sonar to detect chest movements for breathing rate estimation and sleep apnea detection. The proposed system can achieve an error of fewer than ${0.11}$ breaths per minute (bpm) even at a distance of up to ${1}$~\!m. Though the work of~\cite{SleepApeaDetection} utilizes an inaudible frequency range (above $18$~\!kHz), it can still be perceived by animals and infants whose hearing systems are more sensitive to high frequency sounds. 
To deal with this audibility issue (\textit{\textbf{Challenge}}-\ref{clg:audibility}) and handle \textit{\textbf{Challenge}}-\ref{clg:timing estimation trade-off}, BreathJunior improves upon~\cite{SleepApeaDetection} by using white noise for respiration estimation~\cite{NoiseRaspiration}. Before transmitting chirp signals for sensing, it randomizes signal phases in frequency domain and recovers it at a receiver end making the generated sounds less obtrusive. The reported respiration rate error can be as low as 0.4~\!bpm at a distance of 40~\!cm. The work in~\cite{RespTracker} overcomes audibility issue (\textit{\textbf{Challenge}}-\ref{clg:audibility}) by using Zadoff-Chu (ZC) sequence. As we outlined in Section~\ref{sec:waveform for sensing}, the FHSS modulated signal enables fine-grained multipath decomposition and  allows multi-person respiration monitoring simultaneously. The reported error is within 0.6~\!bpm under various test environments in presence of multiple targets at a minimal distance of 10~\!cm. 
Ren et al.~\cite{SleepMonitoring} developed a passive sensing system that can detect breathing rates and sleep-related events from breathing signals. This approach employs high-quality sensors, and reports less than ${0.5}$~\!bpm detection error rates. 

The performance of previous solutions degrades significantly in noisy environments. To combat noise, BreathListener~\cite{BreathListener} extracts breath patterns from energy spectrum density and regenerates clean breath signal via a generative adversarial network. It achieves an average error of 0.11~\!bpm for breathing rate estimation in driving conditions. SpiroSonic~\cite{SpiroSonic} went one step further toward conducting spirometry tests in a regular home setting under various environment noises. It measures a target's chest wall motion via acoustic radar and maps the obtained waveforms to lung function indices. SpiroSonic achieves $5\%$ to $10\%$ monitoring error in clinical studies, allowing reliable out of clinic disease tracking and evaluation. 
%Other respiration sensing systems using acoustic radars include the work in \cite{CoughSensing} that detects coughs, and SpiroSmart \cite{SpiroSmart} and SpiroCall \cite{SpiroCall} that diagnose lung function.
%, EmotionSense \cite{EmotionSense} that identifies psychological states, and StressSense \cite{StressSense} that uncovers stress.

%\begin{table*}[b]
%\caption{Comparison between different keystroke detection systems.}
%\label{tab:keystroke detection}
%\centering
%\begin{tabular}{|c|c|c|c|c|c|}
%\hline
%\textbf{Works} & \textbf{Sensing styles} & \textbf{Waveform design} & \textbf{Key techniques}  & \textbf{Processing latency (ms)} & \textbf{Performance} \\ \hline
%Ubik~\cite{UbiK} & Passive sensing & NA & \makecell{Multipath profiles\\fingerprinting}  & 54.1~\!ms on average & \makecell{Above 95\% detection\\ accuracy on average}\\ \hline
%EchoWrite~\cite{EchoWrite} & Active sensing & 20~\!kHz Tone & \makecell{Doppler profiles\\Bayesian} & $\le$ 200~\!ms & \makecell{94.5\% detection accuracy\\for top 3}\\ \hline
%\cite{SnoopingKeystrokes} & Passive sensing  & NA& \makecell{TDoA ranging\\acoustic profiles} & & 94\% detection accuracy\\ \hline
%\cite{KeystrokeAttack} & Passive sensing & NA & \makecell{TDoA ranging\\inter-connection between\\keystrokes}  & Not reported & 94\% detection accuracy\\ \hline
%\cite{UbiTap} & Passive sensing & NA & \makecell{Acoustic dispersion\\linear regression}  & Not reported & \makecell{Less than 1~\!cm \\localization error, \\98.5\% average\\detection accuracy}\\ \hline
%\end{tabular}
%\end{table*}

\subsubsection{Research Opportunities}
\label{sec:future direction for temporal features}
We envision that research opportunities utilizing temporal features for acoustic sensing lie in two aspects. First, for large-scale deployments, existing approaches need to be made more robust to multipath effects, background noise, interference from ambient sound emitting sources (aroused from \textit{\textbf{Challenge}}-\ref{clg:robust onset detection}) and the heterogeneity of consumer devices (\textit{\textbf{Challenge}}-\ref{clg:calibration}  and \textit{\textbf{Challenge}}-\ref{clg:cross-platform models}). Infrastructure-based methods not only need to minimize initial setup costs, but also need to reduce operating costs, such as (repeated) site surveys and maintenance. Furthermore, solutions that are adaptive to environment changes or hardware upgrades or replacements are attractive. Second, novel applications utilizing accurate timing estimation on acoustic devices can be investigated. One such an example is AcuTe~\cite{AcuTe}, where the authors leveraged the relationship between sound speeds and ambient temperatures for temperature sensing on a single smartphone. Chirp mixing is utilized to estimate the time differences and consequently the propagation speed of sound, which has a linear relationship with temperature. A median measurement accuracy of $0.3^\circ C$ is reported in~\cite{AcuTe}.

\subsection{Solutions Enabled by Channel Characteristics}
\label{sec:channel characteristics-enabled applications}

In this section, we present applications that detect gestures or motions that induce specific characteristics in over-the-air or structure-borne channels. All approaches in this section are device-free. 
%We categorize existing applications into gesture recognition, keystroke detection, touch force detection, authentication, activity recognition, and other interactive sensing systems. 

\subsubsection{Over-the-air Channel Based Approaches}
\label{sec:gesture recognition}
Presence of objects and changes in their shapes and positions in the air affect the channel characteristics and subsequently the reflected acoustic waves. 

\paragraph*{Gesture recognition} 
Gesture recognition aims to understand the expressive meaning of body parts, in particular hands, serving as an interface for humans to interact with smart devices. Previous approaches~\cite{GRSurvey} often rely on dedicated devices or computationally intensive image processing techniques. In contrast, acoustic sensing methods are comparatively much more lightweight. The primary principle of these gesture recognition system is to decipher an appropriate CSI to accurately model acoustic features shadowed by a performed gesture. The key challenge lies in the discovery of this CSI (\textit{\textbf{Challenge}}-\ref{clg:appropriate CIR representation}). To evaluate the performance of a gesture recognition system, one often utilizes the detection or recognition accuracy that is defined by the ratio between the number of correctly recognized gestures and the total performed ones. We hereby present existing works on gesture recognition. 

SoundWave~\cite{SoundWave} utilizes Doppler effect as CSI for gesture recognition. 
SoundWave continuously triggers an inaudible tone and infers gestures by sensing the spectrum of hand-reflected echoes. The key idea is that the reflected acoustic echoes from a moving hand are shifted in the frequency domain compared with the transmitted one. If the hand is moving away, the spectrum of the acoustic echoes is below the transmitted one and vice versa. Combining unidirectional movements allows the recognition of more complex gestures such as flick and quick taps. SoundWave reports recognition accuracy over ${86.67\%}$ in various testbeds. AudioGest~\cite{AudioGest} applies both Doppler effect and multipath profiles for fine-grained gesture recognition. It combines both MD (Doppler frequency estimation) and DD approaches (multipath profiles based linear classifier). AudioGest reports an accuracy of ${96\%}$ in detecting six hand gestures. Another Doppler based gesture recognition system is presented in~\cite{EchoWrite}. This gesture recognition system, called EchoWrite, is an input system based on active acoustic sensing. EchoWrite decomposes the writing gestures into different strokes that exhibit different Doppler profiles estimated from acoustic reflections. Given that an English letter can be generated only by a unique combination of different strokes, a Bayesian inference model is exploited to decipher the written letters via acoustic profiles built on reflections.

UltraGesture~\cite{UltraGesture} utilizes FHSS modulated signals to estimate the CSI in a similar vein with common wireless systems. After then, the estimated CSI is then act as input for a deep learning network for gesture classification. The underlying premise is that using CSI as input can preserve all important information while common methods built on Doppler frequency or multipath profiles would lose many critical information. As a result, UltraGesture can recognize 12 gestures with an accuracy over $97\%$. %\textcolor{green}{To relieve the efforts in constant calibration and improve the recognition accuracy, a data augmentation method is incorporated. }
%UltraGesture~\cite{UltraGesture} integrates both parametric modeling and data-driven approaches for finer gesture recognition. It utilizes acoustic Channel Impulse Response (CSI) to extract specific gesture dependent signatures and than put these signatures into a deep learning framework for recognition. UltraGesture reports over $97\%$ accuracy in recognizing $12$ gestures.
A similar approach can be found in~\cite{PLAGR} that achieves an recognition accuracy of $98.4\%$ with up to 15 gestures.  
Other gesture recognition systems include VSkin~\cite{VSkin} that enables gesture recognition on the back of a device and the work in~\cite{DepthAware} that facilitates depth-aware finger tapping on virtual displays. These two are based on the gesture tracking technology in~\cite{LLAP}. VSkin achieves an accuracy of $99.65\%$ in recognizing tapping events and the work in~\cite{DepthAware} reports a 98.4\% finger tapping detection accuracy. 
 
\paragraph*{Speaker authentication} 
Besides gesture recognition, over-the-air channels can be also exploited for speaker authentication. Commonly used approaches that utilize speech signatures unique to individuals but do not test the ``liveliness" of the speech. Thus, they can fall victims to replay attacks. 
SpeakPrint is a lipreading technology based on acoustic sensing~\cite{SpeakPrint}. It captures how a user speaks by recording mouth and vocal movement through near-ultrasound signals emitted by a  mobile phone at the same time. By features extracted from voice signals and reflected near-ultrasound signals, $100\%$ accuracy is achieved in detecting replay attacks. For user verification, the authors reports an average true positive rate of 99.56\% and a false positive rate of 0.013\%. 

\paragraph*{Novel interactive controls}
The work discussed so far concerns the open air channel between a pair of acoustic speaker and microphone on a smartphone. 
Acoustruments~\cite{Acoustruments} takes a very different approach. It fabricates a tube as an acoustic conduit that connects the transceivers on a commodity smartphone.
The tube has a physical control unit that can move and thus manipulates the properties of received acoustic signals. Through a fingerprinting strategy, the relationship between physical control and received acoustic features can be thus built. 
The authors developed a classification model to recognize different control commands and achieves ${99\%}$ accuracy. Though being innovative and effective, this proposal cannot address \textit{\textbf{Challenge}}-\ref{clg:calibration} and \textit{\textbf{Challenge}}-\ref{clg:cross-platform models}. 

\subsubsection{Structure-borne Channel Based Approaches}
\label{sec:touch force detection}

%\textcolor{green}{As discussed in Section~\ref{sec:acoustic channel property}, the properties of structure-borne channels in a solid medium would be altered when undergoing changes, say external force, . Such properties have been utilized to devise touch interfaces for human-computer interactions. }

Recalled that in Section~\ref{sec:acoustic channel property} that, structure-borne signals in solid medium exhibit two properties, namely, acoustic dispersion and acoustic resonance. The acoustic dispersive property carry range information hence are often used for localization. The acoustic resonance property often reveal rich acoustic features and are more common in interactive applications. We now present respective applications built on structure-borne channel properties. 

%Human-computer interaction with touchable panels can enhance input flexibility. However, prototyping force-sensitive input systems often requires complex circuit designs and hardware configurations.
%Commodity hardware, e.g., smartphones, with such capabilities use dedicated sensors~\cite{3DTouchSensor} that are not pervasively adopted. In contrast, acoustic subsystems are ubiquitous due to the pervasive device support~\cite{MobilePhoneUsage} and thus can enable touch sensing in the wild. A comparison between existing works on force detection can be observed in TABLE~\ref{tab:force detection comparison}. 

UbiTap~\cite{UbiTap} is a keystroke localization system built on a special physical property of structure-borne acoustic signals, called acoustic dispersion. The key observation in UbiTap is that the range information can be precisely obtained by the slope of a straight line shaped between frequency and time in the spectrogram of passively captured keystroke sounds. With this parameteric model, they achieve millimeter-level keystroke localization performance. 

ForcePhone \cite{ForcePhone} estimates applied forces on smartphones with built-in acoustic sensors by exploiting structure-borne sound propagation. It utilizes the fact that emitted acoustic signals from the speaker of a smartphone can cause vibration of the phone body, the intensity of which is inversely proportional to external pressure.
Based on this observation, it builds a close-form parametric model between applied forces and received signal intensity levels. ForcePhone reports a mean square error of 54~\!g at stationary case, which is sufficiently low compared with the maximum 1.5~\!kg sensing range. 

Touch {\&} Active~\cite{TouchForceSensing2} utilizes a data-driven approach that leverages the acoustic resonant property to recognize different touch gestures as well as touch force.
The basic idea is that a certain object has a unique resonant frequency and any touch action can alter it. Such a change manifests itself in the power spectrum of received signals with identifiable features hence can be recognized. This work reports an accuracy of ${99.6\%}$ and ${86.3\%}$ in recognizing five touch gestures and six hand postures on a plastic toy. Meanwhile, it can recognize discrete touch force with per-user recognition accuracy as high as 99.6\%.  

\subsubsection{Research Opportunities}
\label{sec:future direction for channel characteristics}
As mentioned in both \textit{\textbf{Challenge}}-\ref{clg:calibration} and \textit{\textbf{Challenge}}-\ref{clg:cross-platform models}, platform diversity can be particularly detrimental to system performance as it requires constant calibration. To ease the pain, building calibration agnostic parametric models is important. However, as acknowledged in \textit{\textbf{Challenge}}-\ref{clg:appropriate CIR representation}, it requires sophisticated domain knowledge and is remarkably challenging. Alternatively, one can leverage deep learning techniques that have shown promising results in  generalizability. Techniques such as adversarial training~\cite{DomainAdaptationSurvey}  allow a model to generalize well on even unseen data, while few-shot learning  techniques~\cite{FewshotLearning,RFComm} only require limited labels to quickly adapt on target environments. Some early attempts in this direction has been made. For example, the authors in~\cite{RFNet,MetaSense} employ meta learning~\cite{MetaLearning}, a kind of shot learning technique, to facilitate cross-device mobile sensing with only one or two data instances. We envision that more sophisticated deep learning approaches will be incorporated in acoustic sensing based on channel characteristics to address device, environment or subject diversity. 

\section{Future Directions}
\label{sec:future direction}
In the previous sections, we have highlighted unique research opportunities in different categories of acoustic sensing applications. In this section, we discuss a few research directions that we believe are under-investigated or are emerging in acoustic sensing in the future. 
\subsection{Privacy and Security Threats}
The ever increasing applications of acoustic sensing also bring about many privacy and security issues. 
For instance, a recent study on acoustic sensing found that a recording system can act as an acoustic mixer~\cite{BackDoor}, making it possible to detect ultrasonic signals above ${24}$~\!kHz on commodity mobile devices with no more than ${48}$~\!kHz sampling rate. This phenomenon, caused by the non-linearity of acoustic sensors, has been exploited in jamming and communication~\cite{BackDoor}. The interesting finding can lead to innovative applications. However, such a technology also poses security threats on smart IoT devices with on-board microphones such as Google Home~\cite{GoogleHome} and Amazon Echo~\cite{AmazonEcho}. It allows synthesizing audio signals inaudible to humans to manipulate such devices~\cite{DolphinAttack, LipRead} and thus open doors to malicious attacks.  Additionally, existing work on keystroke detection~\cite{KeystrokeAttack,SnoopingKeystrokes,UbiK} can pose privacy threats if enabled by malicious attackers with the knowledge of legitimate users. 
Therefore, techniques to detect and defend against such attacks constitute an important area of investigation for acoustic sensing.

\subsection{Repurposing Other Sensors for Acoustic Sensing}

%XXX: need more references. besides RADAR or LIDAR any other sensors? vibration sensor? 
In addition to use acoustic sensors for non-conventional purposes such as touch sensing and gesture tracking, it is possible to use other sensors to capture acoustic signals. 
%For instance, the authors of~\cite{SLAMEF} leverages high-sensitivity microphones to detect magnetic signals from electric wires as these magnetic signals can influence the current flowing a circuit, which can be detected by a microphone. The authors further find that such influence is location-dependent and therefore build a simultaneous localization and mapping system.  
For instance, the authors from~\cite{UWHear,WaveEar} and \cite{SpyWithLidar} exploit RF-radar and Lidar sensor to hack audio content, respectively. The key observation is that acoustic signals originate from  vibrations, which can be detected through sensors that measure displacement. A main benefit of cross-technology sensing is that these non-dedicated acoustic sensors are immune to background acoustic noise and thus provide high SNR signals as long as the sampling rate is adequate. 
%XXX: why ? hardware availability? restricted sensing range??
%Conversely, other events that cause object vibration can also be detected by acoustic microphones. One such example can be found in \cite{EarSense} where the authors repurpose earables as microphones for touchable interaction. 
Therefore, an interesting research direction is to incorporate data from multiple sensing modalities to enable sophisticated sensing tasks or novel applications in adverse environments. 
%We envision that such innovative utilization of sensors would flourish in the near future and hence enable many more interesting applications.
%
%XXX: accelerometer? 
%[CC]: may not be sensitive enough, and the sampling rate is insufficient
\subsection{Earable Computing}
With the prevalence of earable devices such as headsets and earbuds, many interesting new applications arise. Almost all earables come with on-board microphones and speakers, making them suitable for acoustic sensing. Additionally, newer devices such as Apple Airpods pro feature IMU sensors and force sensors. Fusing the measurements from multiple sensors opens up new areas of research. In~\cite{EarableComputing}, Choudhury provides an excellent overview of earable computing research opportunities. Notably, smart earables can be used to enhance human auditory perception in the form of hearing aids~\cite{Hearable}, binaural sound localization~\cite{BinauralLocalization}, and 3D spatial sound spatialization~\cite{EarAR}. By characterizing the channel responses of inner ear canals, one can extract useful information for user authentication~\cite{EarEcho} or diagnosing ear infections~\cite{EarFluidDetection}. From structure-borne signals and IMU sensor data, chewing and drinking activities can be monitored~\cite{EarRecognition}. Structure-borne signals can also be used to enhance the recognition of one's speech when air-borne signals are of poor quality in noisy environments. To realize these applications, in addition to common performance measures, it is important to minimize computation and communication overheads due to the limited form factor and battery lifetime of earable devices. 
\subsection{Benchmark Datasets and Platforms}
To date, there is no open datasets or common platforms to benchmark acoustic sensing solutions. The lack of open datasets can be in part attributed to the need to design customized waveforms in specific applications. As discussed in Section~\ref{sec:acoustic hardware}, hardware diversity among COTS devices render solutions tailored to one platform not suitable for others. Absence of common platforms also means that researchers often need to start from scratch and spend much time handling platform-dependent details such as programming languages and development frameworks. To overcome these problems, the acoustic sensing community can learn from successful practices in other communities. For instance, CRAWDAD is a wireless network data resource maintained by researchers at Dartmouth University with contributors from all of the world~\cite{CRAWDAD}. ORBIT~\cite{ORBIT} is a two-tier wireless network emulator/field trial designed to achieve reproducible experimentation, while also supporting realistic evaluation of protocols and applications. It is available for remote or on-site access by academic researchers both in the U.S. and internationally. GNURadio and USRP platforms provide unified hardware platforms and basic processing blocks for radio signal processing and propel research in advanced wireless technologies~\cite{GNURadio,USRP}. 

Recognizing the need of common platforms, the work in~\cite{MobileSensingPlatform} and~\cite{MobileSense} are among the first to build general sensing platforms on commodity mobile devices. Later on, the authors in \cite{LibAS} released a cross-platform support for ubiquitous acoustic sensing. Cai et al. develop ASDP, the first Acoustic Software Defined Platform, which encompasses several customized acoustic modules running on a ubiquitous computing board, supported by a dedicated software framework~\cite{ASDP}. Unfortunately, these platforms have not yet gained much attraction in the community.  We believe significant efforts are needed toward data sharing and advocate the adoption of common platforms for evaluation. 
\section{Conclusion}
\label{sec:conclusion}
This paper presented a systematic survey on acoustic sensing. We provided in-depth discussions on the unique challenges posed by acoustic hardware, systems and channels as well as techniques to mitigate them. We organized related research in a bottom-up manner from the physical layer to application-layer solutions to expose researchers and developers with a systematic view of what a full system entails. Discussions on research opportunities and future directions aimed to spawn further efforts in this area and encouraged the research community to take an concerned effort to transform research outcomes to real-world practices and products.

%In this paper, we presented a comprehensive survey on acoustic sensing. And we developed a layered architecture for acoustic sensing systems. This architecture encompasses three layers, namely, application layer, processing layer, and physical layer. In the application layer, we classify existing work into three categories: namely context-aware application, human-computer interface, and aerial acoustic communication. We then discuss the basic idea and the performance behind each category. 
%In the processing layer, we comprehensively analyze different sensing approaches. In the physical layer, we present fundamental design considerations in details.

%Despite tremendous developments in acoustic sensing, there are still many technological challenges need further investigation, i.e., user configuration, multipath effect, sampling frequency offset, heterogeneity, and system delay. We believe that solutions to these challenges will not only improve system performance but also lead to a rise in many exciting applications. At the end of the survey, we introduced several untapped and attractive research topics that are worth of further investigations. By the timely and thorough review of existing work, this survey may serve as guidelines and encourage more research efforts into acoustic sensing.

\bibliographystyle{abbrv}
\bibliography{temp}

\end{document}